\documentclass[12pt]{article}
\usepackage{graphicx}
\usepackage{amssymb}
\begin{document} 

\begin{titlepage}
	\rightline{}
	
	
	\vskip 2cm 
	\begin{center}
		\Large{{\bf On black hole thermalization,\\
		D0 brane dynamics, and emergent spacetime}}
	\end{center}
	
	\vskip 2cm 
	\begin{center}
		{Paul Riggins\footnote{\texttt{priggins@hmc.edu}}\ \ \ and\ \ \ Vatche Sahakian\footnote{\texttt{sahakian@hmc.edu}}}\\
	\end{center}
	\vskip 12pt 
	\centerline{\sl Harvey Mudd College} 
	\centerline{\sl Physics Department, 241 Platt Blvd.}
	\centerline{\sl Claremont CA 91711 USA}
	
	\vskip 1cm 
	\begin{abstract}
		When matter falls past the horizon of a large black hole, the expectation from string theory is that the configuration thermalizes and the information in the probe is rather quickly scrambled away. The traditional view of a classical unique spacetime near a black hole horizon conflicts with this picture. The question then arises as to what spacetime does the probe actually see as it crosses a horizon, and how does the background geometry imprint its signature onto the thermal properties of the probe. In this work, we explore these questions through an extensive series of numerical simulations of D0 branes. We determine that the D0 branes quickly settle into an incompressible symmetric state -- thermalized within a few oscillations through a process driven entirely by internal non-linear dynamics. Surprisingly, thermal background fluctuations play no role in this mechanism. Signatures of the background fields in this thermal state arise either through fluxes, {\em i.e.} black hole hair; or if the probe expands to the size of the horizon -- which we see evidence of. We determine simple scaling relations for the D0 branes' equilibrium size, time to thermalize, lifetime, and temperature in terms of their number, initial energy, and the background fields. Our results are consistent with the conjecture that black holes are the fastest scramblers as seen by Matrix theory.
	\end{abstract}
\end{titlepage}

\newpage \setcounter{page}{1} 
\section{Introduction and results}
\label{sub:intro}

The correspondence between strongly coupled gauge theories and quantum gravity~\cite{Maldacena:1997re}-\cite{Banks:1996vh} suggests that gravitation and spacetime may be viewed as emergent structures -- emergent from within the intricate non-linear interactions of non-gravitational quantum field theories. One aspect of this correspondence that remains particularly challenging to understand has to do with a thermodynamics theme that seems to underly gravitational dynamics~\cite{Witten:1998qj,Maldacena:2001kr}. 

A representative illustration of this problem arises in the following process. A stringy -- possibly Planckian -- probe falls into a black hole. The expectation from the gravity-gauge theory correspondence is that the probe gets thermalized as it flies through the spacetime region near the horizon of the black hole~\cite{Balasubramanian:2011ur}. What underlies this mechanism of thermalization on the gravity side? More precisely, what is the spacetime that the probe actually sees as it crosses the horizon -- if not the smooth, traditional, no-hair region that ends with a pathological singularity. Furthermore, there have been suggestions that this thermalization process is an unusual one, characterized by a fast scrambling of the information in the probe~\cite{Sekino:2008he}. The situation also ties in with the black hole information paradox that attempts to account for such scrambled information~\cite{Mathur:2010kx,Mathur:2011uj,Mathur:2012ux}. In a slightly different language, we want to find out why does the traditional picture of the spacetime near a black hole horizon fail to fully capture horizon physics -- even when the curvature scales at the horizon are very small.

In this work, we investigate these questions using the strongly coupled $0+1$ dimensional gauge theory that describes D0 brane dynamics~\cite{Banks:1996nn,Douglas:1996yp,Banks:1996vh}. We want to quantify the thermalization process on the gauge theory side using numerical simulations; and identify the role played by background fields that the D0 branes are immersed in. If a spacetime -- be it that of a traditional black hole or one of the many fuzzball geometries~\cite{Mathur:2005zp,Mathur:2008nj} -- is to thermalize a probe, in what way does this spacetime fix the thermodynamic attributes of this thermalization process? For example, if a probe is to get scrambled into a configuration of temperature determined by the size of a black hole horizon, we want to find out how the geometrical information about the size of the horizon gets eventually encoded into the temperature attribute of the scrambled probe. Through this, we can start addressing the difficult question as to how one determines the emergent spacetime that an in-falling probe actually experiences -- assuming the traditional no-hair geometry somehow falters in the vicinity of a horizon.

The numerical simulation of $0+1$ dimensional $U(N)$ Super Yang-Mills theory at strong coupling is a difficult one -- even without the inclusion of supersymmetry --- since it is particularly computationally intensive~\cite{Kaplan:2002wv,Catterall:2007fp,Anagnostopoulos:2007fw,Hanada:2010rg}. Larger values of $N$ provide for numerical stability at the expense of time. And a thermodynamic treatment requires a large ensemble of simulations. We overcome these problems by a series of efficient physical and technical tricks that we develop; and through the use of recent technological advances in parallel processing. The result is a framework of D0 brane dynamics exploration that can be done in real-time, pocking and tweaking the parameters as the dynamics evolves so as to develop physical intuition about this very rich system.

We track the time evolution of a probe consisting of $N$ D0 branes in various background field configurations -- in their center of mass frame. The D0 brane coordinates are represented by $N\times N$ hermitian traceless matrices $x^\mu$, nine in total, one for every spatial direction. The eigenvalues of these matrices may be interpreted as the positions of the individual D0 branes. We define {\em the extent} of the probe in a subspace of the nine space directions as
\begin{equation}
	\rho_{x^\mu}^2 = {{\mbox{\bf Tr} \left( x^\mu x^\mu \right)}}
\end{equation}
where $\mu$ is one representative index within the subspace of interest (no sum over $\mu$). The actual size of the probe is defined as~\cite{Myers:1999ps}
\begin{equation}\label{eq:sizedef}
	R \equiv \sqrt{\frac{\mbox{\bf Tr} \left( x^\mu x^\mu\right)}{N}}
\end{equation}
for $\mu$ summed over $1\ldots 9$.
In all cases, we start the D0 branes in a static two-sphere configuration in three of the nine space directions
\begin{equation}
	x^i = s\, \sigma^i\ \ \ ,\ \ \ \left[ \sigma^i, \sigma^j \right] = i \varepsilon^{ijk} \sigma^k\ \ \ \mbox{For $i,j,k=1,2,3$}
\end{equation}
where the $\sigma^i$'s are the Pauli matrices in an $N\times N$ representation, and where $s$ is a tunable parameter that fixes the initial size. This initial size is  then given by
\begin{equation}\label{eq:Rsize}
	R = s\, \sqrt{2}\sqrt{N^2-1} \simeq s\,\sqrt{2} N
\end{equation}
for $N\gg 1$.
Tuning $s$ also corresponds to adjusting the initial energy of the setup: smaller $s$ corresponds to smaller initial energy. We refer to the six directions transverse to the initial spherical configuration as the transverse directions; and we refer to the remaining three as probe directions. Matrices in the three probe directions are denoted by $x^i$ with $i=1,2,3$; while those in the transverse directions are denoted by $y^a$ with $a=1\ldots 6$, initially set to zero. The momenta matrices in the probe directions are denoted by $p^i$, while those in the transverse directions are labeled $q^a$. We also add small random gaussian fluctuations to the initial positions of the D0 branes -- off the spherical shape of the probe. This is so as to avoid starting on top of a saddle point in the potential, ending up exploring non-generic regions of the phase space. This initial randomization of the positions can be viewed as accounting for the effect of initial quantum or thermal fluctuations~\cite{Asplund:2011qj}. 

\begin{figure}
	\begin{center}
		\includegraphics[width=7in]{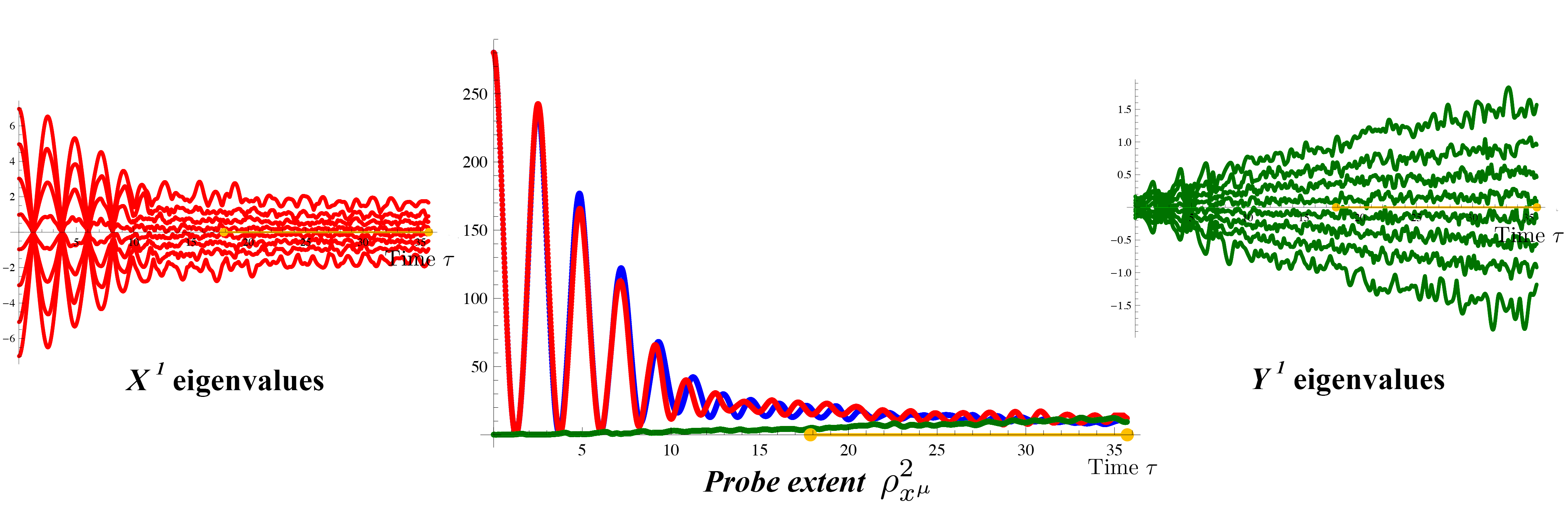} 
	\end{center}
	\caption{The evolution of D0 branes in zero but fluctuating background fields. In this case, we have $15$ D0 branes whose initial energy is much larger than the energy in the fluctuating background fields. The horizontal axis is time $\tau$ in all three graphs. The smaller two graphs are plots of a selection of eigenvalues of the D0 branes: in one of the three probe directions on the left, and in one of the six transverse directions on the right. In the middle, the graph shows the extent of the probe: in red and blue for two probe directions, and in green for a transverse direction. All variables are dimensionless, as defined in the main text.}\label{fig:energetic}
\end{figure}

In this work, we study over $500$ simulations, all qualified with a numerical error of the order of $10\%$. Figure~\ref{fig:energetic} is a stereotypical time 
evolution when the initial conditions are perturbed slightly. Without initial perturbations, we find oscillatory evolution with no thermalization. Throughout, lengths are measured in units of $\ell$, and energies and temperatures in units of $1/\ell$, where 
\begin{equation}
	\ell \equiv (2\pi)^{2/3} \frac{l_s}{g_s^{1/3}} \gg l_s\ ,
\end{equation}
with the string coupling $g_s\ll 1$ and the string length given by $l_s$. The spherical configuration oscillates to a smaller extent in the probe directions; at the same time, its initial zero extent in the transverse directions expands to form a uniform fuzzy ball in all nine space directions. In a short time -- within a few oscillations -- the eigenvalue spectrum becomes thermal. The effective Yang-Mills coupling is mostly large -- {\em i.e.} the thermalization phenomenon is a strong coupling effect\footnote{By this we mean that the non-linear terms in the Hamiltonian evolution equations play a central role. The dynamics is classical chaotic only if these terms are initially large, and we find that, if the initial energy is tuned such that the non-linearities are weak, there is no thermalization.}. There are various interesting attributes of this process that we discover as we vary the background parameters. We explore three qualitatively different scenarios. First, we consider zero but thermally fluctuating background fields. This corresponds to exploring the effects of a dilute gas of massless supergravity fields onto the probe. In this analysis, we include the effects of thermal back-reaction -- a central but delicate mechanism to the process of thermalization in traditional thermodynamic systems. We consider only a fluctuating background metric and D2 brane flux for simplicity. Next, we consider non-zero background fields with negligible fluctuations and study the effects of non-zero gravity and D2 brane flux on D0 probe thermalization. Finally, we model the problem of the probe falling into a large Schwarzschild black hole and determine the effect of the in-fall on the thermalization of the probe. We summarize the results in the upcoming three sections. 

\subsection{Results: zero backgrounds}

\begin{figure}
	\begin{center}
		\includegraphics[width=7in]{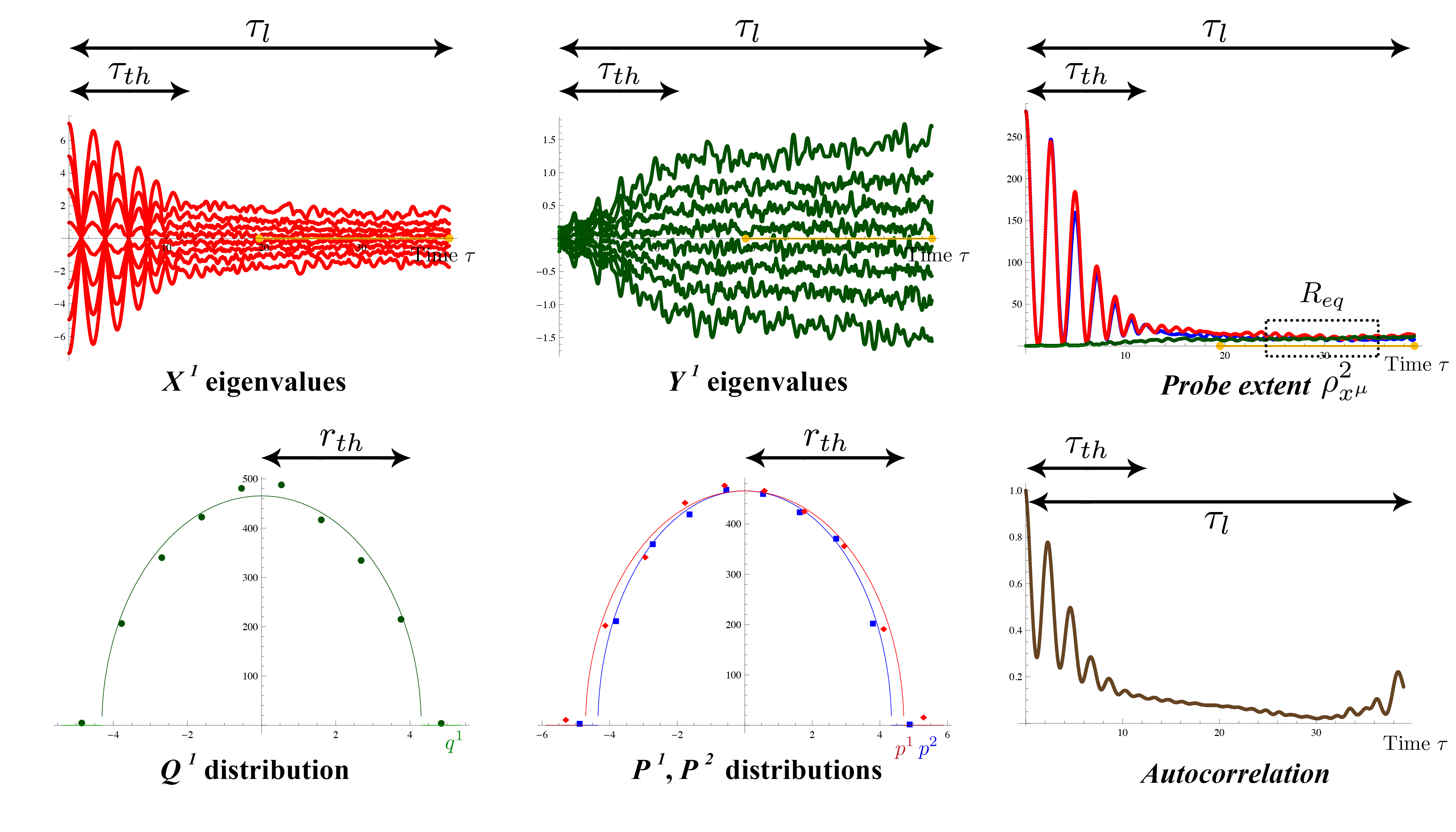} 
	\end{center}
	\caption{A sample simulation, along with a depiction of the output parameters: $\tau_{th}$ for the time to thermalize, $\tau_l$ for lifetime, $t_p=r_{th}^2$ for probe temperature, and $R_{eq}=\sqrt{9}\times\rho_{x^\mu}/\sqrt{N}$ for final probe size. For this simulation, the input parameters were: $N=15$, $s=1$, with no background fluctuations. In the probe extent and distribution graphs, red represent the probe direction $x^1$, blue is for the probe direction $x^2$, and green is for the transverse direction $y^1$.}\label{fig:effectofs}
\end{figure}

We start with zero background fields -- with or without thermal fluctuations of the supergravity fields. Even when the background fluctuations are zero, we introduce a level of randomization into the problem by perturbing the initial spherical configuration of the D0 branes at around $10\%-50\%$ level. We can vary $N$, the number of D0 branes in the probe, and the initial size of the probe sphere -- through the parameter $s$ which also tunes the initial energy. Figure~\ref{fig:effectofs} shows the data we collect from a simulation for $s=1$ and $N=15$ D0 branes. In this case, we have no background field fluctuations. These six plots are generated for every simulation, and we use this example to outline the typical analysis we perform for any simulation. The top three plots in the figure  
show the eigenvalues and extent of some of the position matrices. The lower three plots attempt to quantify the thermalization phenomenon. The two leftmost graphs on the bottom are histograms of the eigenvalues of momenta matrices, computed for the last half of the simulation timeline where the probe has settled into a potentially thermal configuration. A semicircle pattern is indicative of thermalization, with the radius of the semicircle giving us a measure of the probe's temperature $t_p=r_{th}^2$)~\cite{Asplund:2011qj}. The rightmost plot on the bottom is a graph of the autocorrelation function
\begin{equation}
	\mbox{\bf Re} \left< \mathcal{O}^\dagger(0) \mathcal{O}(\tau)\right>
\end{equation}
as a function of time $\tau$; and the $\mathcal{O}$ operator is given by
$\mathcal{O}=\mbox{Tr}\left( X^1+i Y^1 \right)^2$. This measures the autocorrelation of a gauge invariant operator over time: for a thermalizing state, the function should decay exponential over time. We see from the example depicted in the figure that the probe does indeed thermalize. From these graphs, we extract the timescale of thermalization $\tau_{th}$ and the probe temperature $t_p$ as shown. Eventually, our simulation invariably breaks down. We track this by looking at the constraint equation
\begin{equation}
	\left[ x^i, p_i\right] + \left[ y^a, q_a \right] = 0
\end{equation}
which needs to be satisfied for a consistent dynamical system. This break-down is seen to occur often (but not always) because the size of the probe suddenly explodes. It signals the accumulation of significant numerical errors by the evolution algorithm. For zero background fields, there is a flat direction in the system when all nine matrices become mutually commuting. Then the configuration can expand unbounded with no cost of energy -- the larger in size the better from entropic considerations. It is likely that the numerical errors that we see in the simulations are due to this instability some of the time. Hence, we associate a lifetime with the probe, denoted by $\tau_l$, the time it takes for the probe to de-stabilize and fly apart, measured from the start of the simulation as shown in the Figure. We are uncertain whether this quantity $\tau_l$ is a physical one, or a pathology of numerical simulations. Finally, from a blow-up of the size plot, we can read off the equilibrium size of the probe $R_{eq}$. 

Hence, each simulation gives us four outputs: $\tau_{th}$, $\tau_l$, $t_p$, and $R_{eq}$. We want to determine the dependence of these quantities on three input parameters: the initial size of the spherical probe tuned by $s$ -- equivalent to keeping track of the energy; the number of D0 branes $N$; and when we turn on thermal background field fluctuations, the temperature $t$ of the background gas.

\begin{figure}
	\begin{center}
		\includegraphics[width=7in]{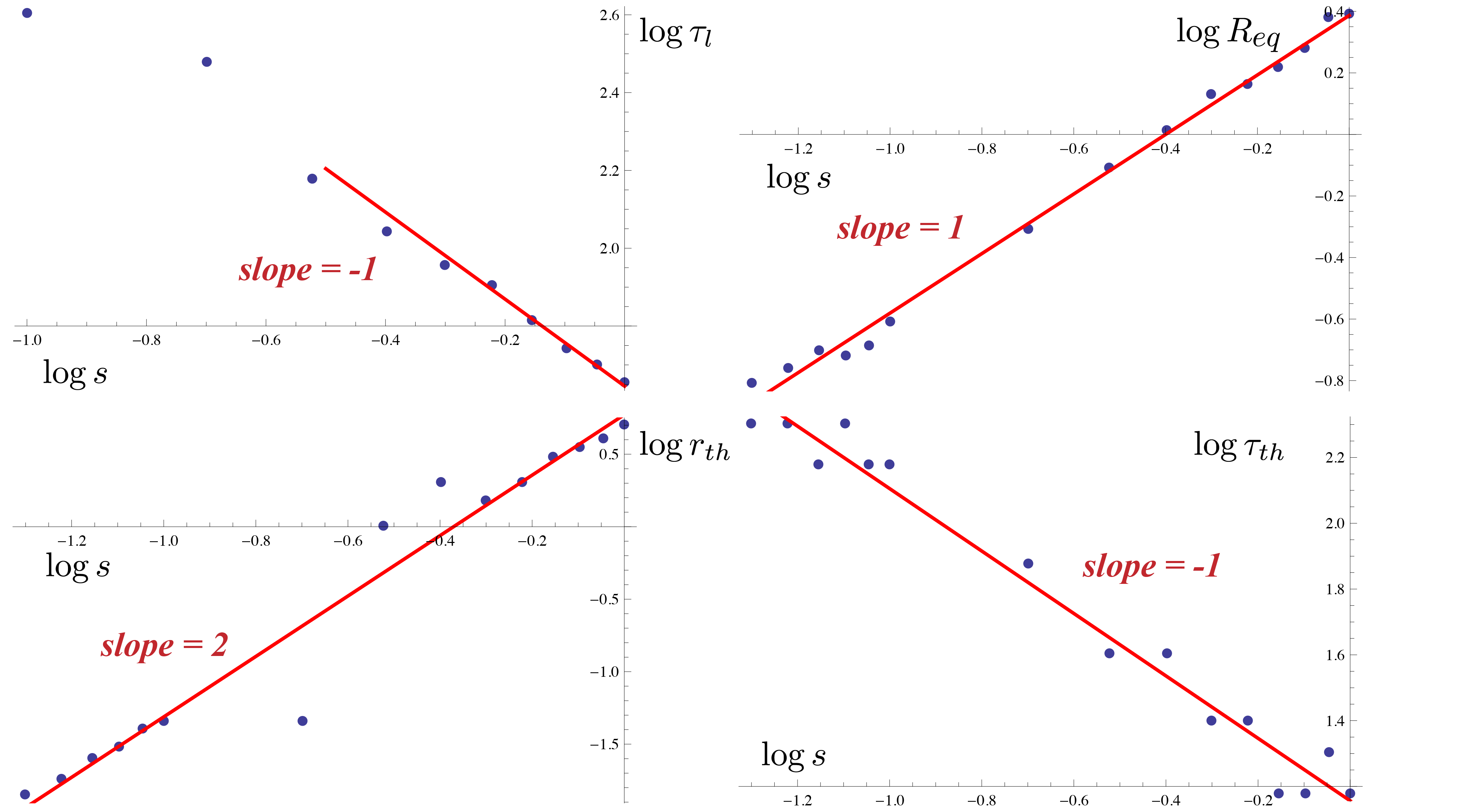} 
	\end{center}
	\caption{Results from a collection of simulations showing the output parameters $\tau_{th}$ (thermalization time), $\tau_l$ (lifetime), $R_{eq}$ (final equilibrium size), and $t_p=r_{th}^2$ (probe temperature) as a function of the input parameter $s$. For all simulations, $N=15$, and there are no background fluctuations. All slopes are within the indicated nearest rational values that is allowed by the $10\%$ error of the simulations.}\label{fig:sscan}
\end{figure}

\begin{figure}
	\begin{center}
		\includegraphics[width=7in]{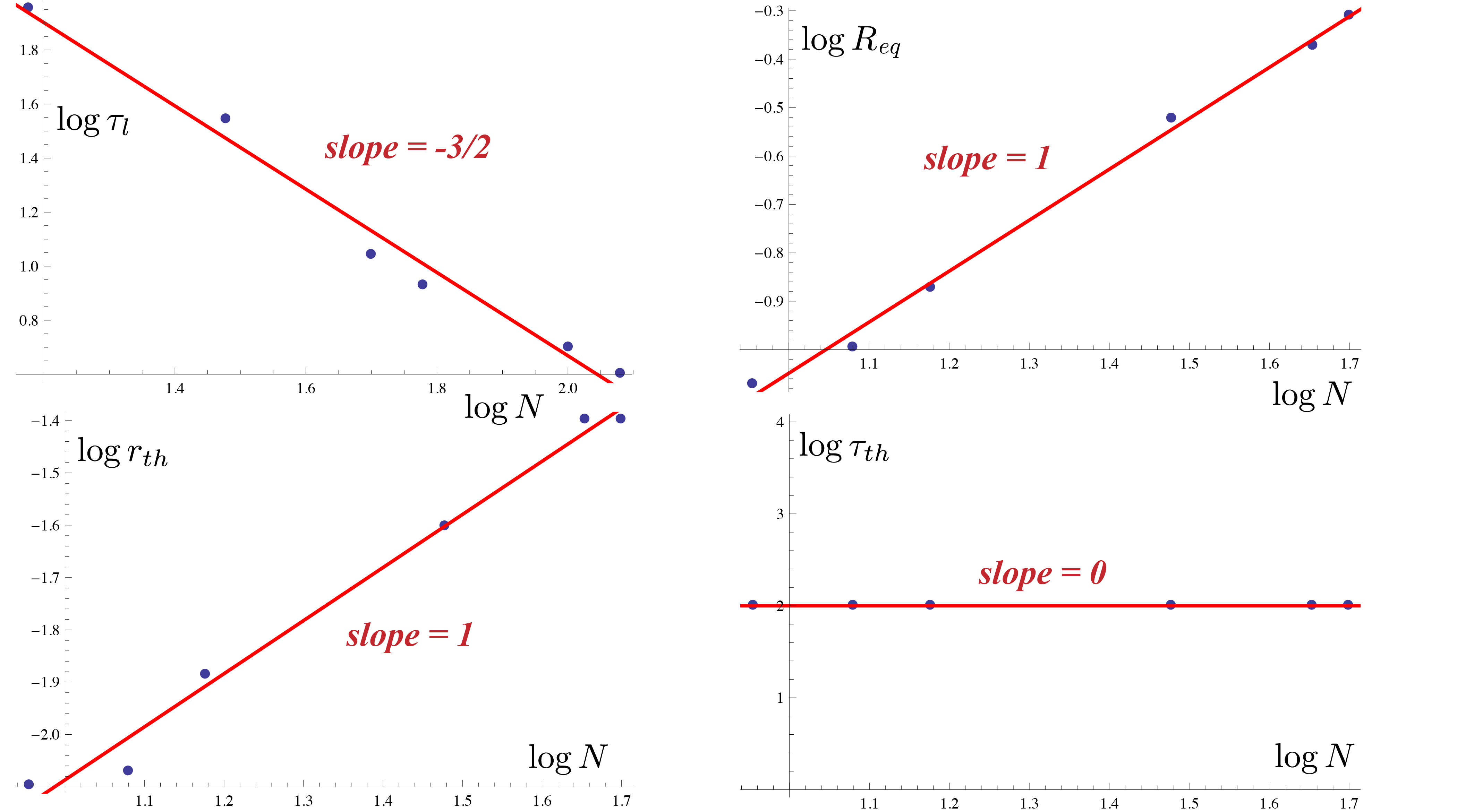} 
	\end{center}
	\caption{Results from a collection of simulations showing the output parameters $\tau_{th}$ (thermalization time), $\tau_l$ (lifetime), $R_{eq}$ (final equilibrium size), and $t_p=r_{th}^2$ (probe temperature) as a function of the input parameter $N$, the number of D0 branes. For all simulations except the ones used to gauge the lifetime $\tau_l$, $s=0.03$, with no background fluctuations. For the graph of the lifetime dependence on $N$, another set of simulations were used due to technical limitations with storing large matrices for long simulations. For the latter case, $s=0.5$ with much shorter simulation times, with no background fluctuations, and larger statistical errors. All slopes are within the indicated nearest rational values that is allowed by the $10\%$ error of the simulations.}\label{fig:Nscan}
\end{figure}

\begin{figure}
	\begin{center}
		\includegraphics[width=7in]{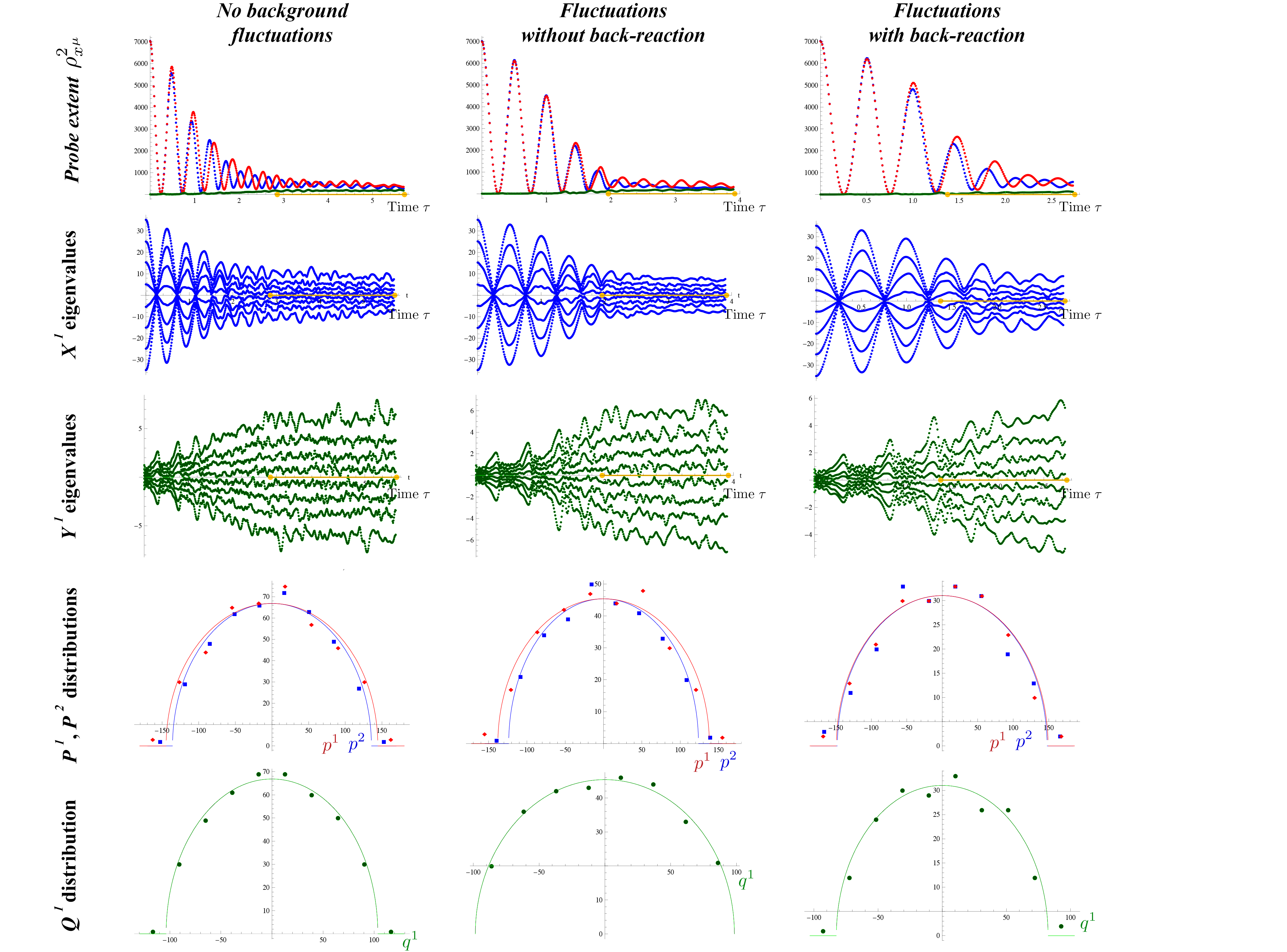} 
	\end{center}
	\caption{The effect of background fluctuations. For all graphs, we have $t=1$, $N=15$, $s=5$, and $g_s=10^{-4}$. We see that background fluctuations do not effect the thermal properties of the probe, except for its lifetime. The size of the fluctuations is determined by the temperature $t$, as well as the string coupling $g_s$, as described in detail in the main text.}\label{fig:temp}
\end{figure}

Figure~\ref{fig:sscan} shows a compilation of the results from a series of simulations which fix all parameters and scan over $s$.
Figure~\ref{fig:Nscan} shows a compilation of the results from simulations which fix all parameters and scan over $N$ instead, the number of D0 branes. For all of these simulations, fluctuations of the background fields are turned off.
If the background fields are made to thermally fluctuate with temperature $t$, there is no signature of this temperature in the probe's thermalization dynamics as depicted in Figure~\ref{fig:temp}. We have scanned over several orders of magnitude of background temperatures $t$ within the regime of validity of our formalism, and considered the delicate effect of thermal back-reaction -- with no change in the conclusion. 
Putting all these results together, we find the following scaling relation for the equilibrium size $R_{eq}$
\begin{equation}\label{eq:relation1}
	R_{eq}\propto s\,N\ .
\end{equation}
For the thermalization time, we find
\begin{equation}\label{eq:scramblingtime}
	\tau_{th} \propto \frac{1}{s} \propto \frac{N}{R_{eq}}\ .
\end{equation}
And for the temperature of the probe $t_p$, we find
\begin{equation}\label{eq:scramblingtemperature}
	r_{th}^2 = t_p \propto s^{4} N^2 \propto \frac{R_{eq}^4}{N^2}\ .
\end{equation}
As for the lifetime of the probe, we find
\begin{equation}\label{eq:relation4}
	\tau_{l} \propto \frac{1}{s\,N^{3/2}} \propto \frac{1}{R_{eq}\sqrt{N}}\ .
\end{equation}
Irrespective of the details, the D0 brane probe seems to eventually disintegrate in a time $\tau_l$. The thermalization time $\tau_{th}$ is of particular interest given the proposal of~\cite{Sekino:2008he,HaydenPreskill:2007} suggesting that black holes are highly efficient scramblers -- with their scrambling timescale proportional to the logarithm of the number of degrees of freedom. We will comment on the implications of our results with regards to this proposal in the Conclusion section at the end.

If the probe is too small and the background gas temperature $t$ is too large, the probe's thermalization is disrupted and the evolution follows deterministic oscillations. 
We cannot determine whether this is a physical effect or a numerical pathology since these simulations lie at the edge of the regime of validity of our formalism. We also consider `on-shell background fluctuations': that is, background field fluctuations that satisfy the Laplace equation at the center of mass of the probe. This corresponds to a scenario where there is no matter sourcing the background fields at the probe's location. Once again, thermalization is disrupted and we are unable to determine whether the effect is physical or numerical. Either way, these scenarios are uninteresting since they do not lead to thermalization.

All four of these relations~(\ref{eq:relation1}) to~(\ref{eq:relation4}) are insensitive to the temperature of the background $t$! A thermal bath of massless supergravity fields seems to play no role in the thermalization of the probe. 

The conclusive theme of the analysis can be summarized as follows: the thermalization of a probe of D0 branes is an internal process arising from the non-linear D0 brane interactions and strong Yang-Mills coupling. A quartic term in the D0 brane matrix coordinates in the Hamiltonian of the system underlies this thermalization phenomenon. The result depends  on the number of D0 branes and their equilibrium size. Fluctuations in the background fields do not play a relevant role in this process, a fact that is rather counter-intuitive as compared to more traditional statistical mechanical systems. There is however a role played by the fluctuating fields in destabilizing the configuration at high enough temperatures. This regime is however at the edge of the regime of validity of our analysis. 

The size of the probe may be sensitive to large non-zero background fields, as opposed to small fluctuating ones. We will investigate this case next.

\subsection{Results: quasi-static non-zero backgrounds}

We next consider non-zero background fields with negligible thermal fluctuations. We turn on the effect of the background metric through tidal gravitational forces acting on the probe in its center of mass frame; and we also turn on background D2 brane flux. These fields may be arising from a large number of background D-branes that our probe is inserted into. If the background is to represent a black hole, we may expect D2 brane flux from previous investigations of Matrix black holes~\cite{Horowitz:1997fr,Banks:1997tn,Banks:1997hz}. 

\begin{figure}
	\begin{center}
		\includegraphics[width=7in]{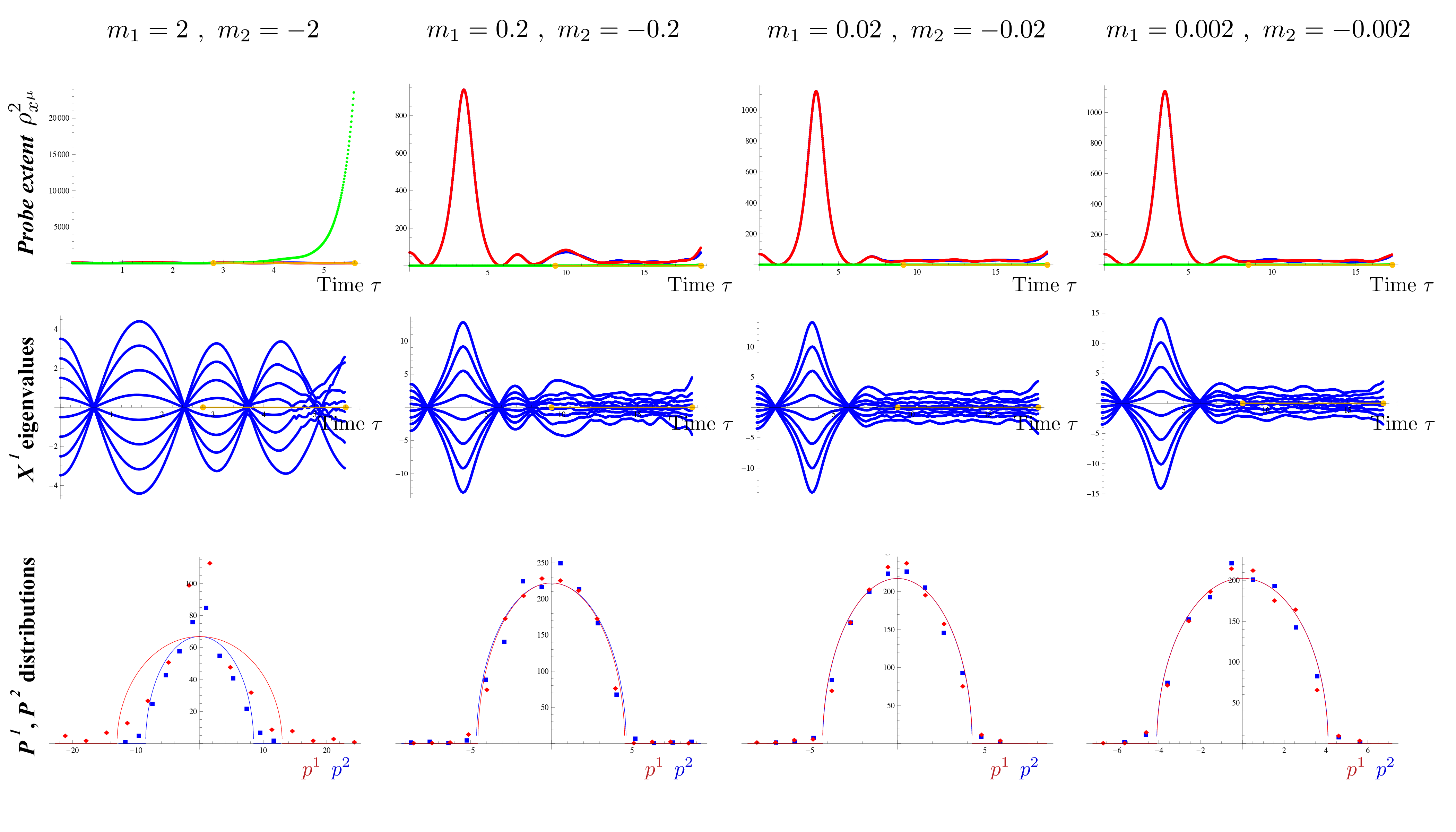} 
	\end{center}
	\caption{The effect of scanning over the gravitational tidal force parameters $m_1$ and $m_2$. A selection of simulation are shown, but many more were analyzed with similar qualitative conclusions. For all simulations in this figure, $s=0.5$, $N=15$, with no background fluctuations. At around $|m_1|=|m_2|=2$, we see the start of the probe size competing with the length scale associated with the background.}\label{fig:mscan}
\end{figure}

\begin{figure}
	\begin{center}
		\includegraphics[width=6in]{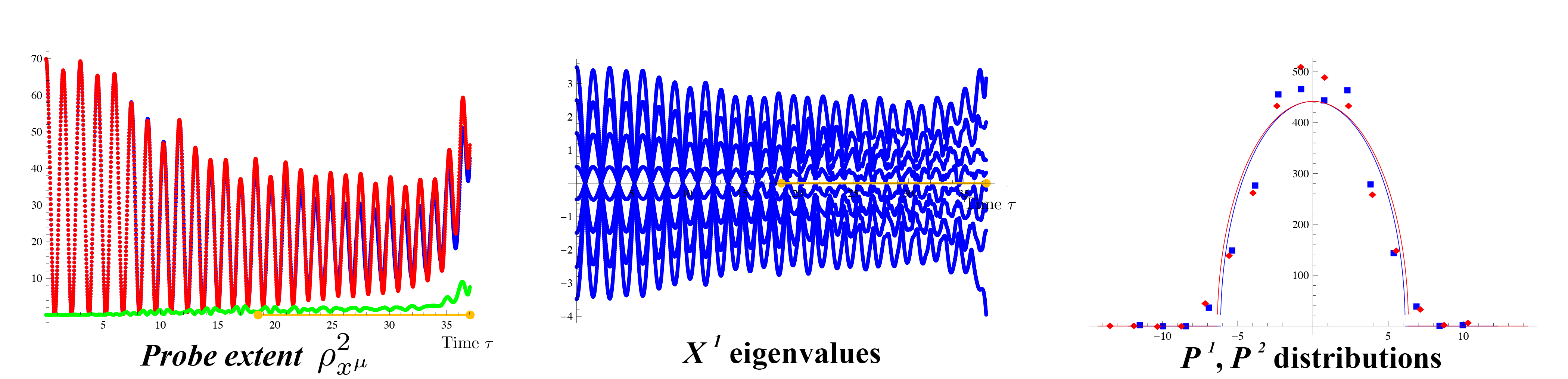} 
	\end{center}
	\caption{When the probe size becomes bigger that the length scale characterizing the background metric, we see a ringing effect and the break-down of thermalization. The validity of our simulations is also breaking down in this region of the parameter space. For this simulation, we have $m_1=m_2=$, $n=-0.05$, $s=0.5$, $N=15$, with no background fluctuations.}\label{fig:ringing}
\end{figure}

We track the dimensionless curvature scales in the three and six dimensional subspaces through the two parameters $m_1$ and $m_2$ respectively. As Figure~\ref{fig:mscan} shows, we 
find no effect of $m_1$ and $m_2$ on lifetime, thermalization time, temperature, or the final size of the probe. That is unless the length scale associated with the $m$'s becomes small enough to compete with the size of the probe. This is akin to the gravitational Gregory-Laflamme phenomenon, and syncs well with the gravitational analogue of this scenario presented in~\cite{Murugan:2006sn}. At this point, we get a ringing behavior as shown in Figure~\ref{fig:ringing}. Thermalization is disrupted and we do see a complex correlation 
between the background's and the probe's thermal characteristics. Unfortunately, this regime tests the bounds of validity of our simulation and cannot be reliably quantified yet. 

\begin{figure}
	\begin{center}
		\includegraphics[width=7in]{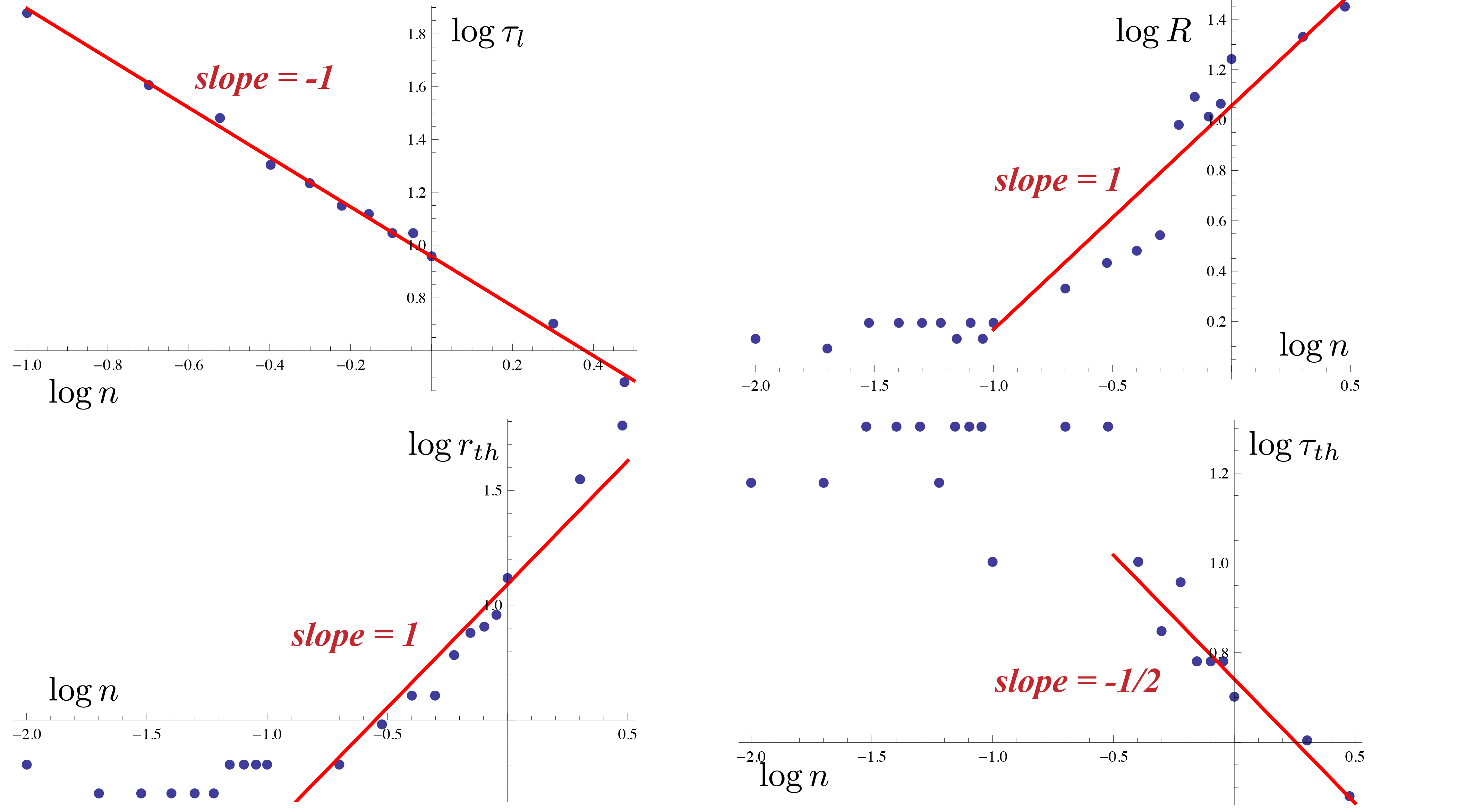} 
	\end{center}
	\caption{Results from a collection of simulations showing the output parameters $\tau_{th}$ (thermalization time), $\tau_l$ (lifetime), $R_{eq}$ (final equilibrium size), and $t_p=r_{th}^2$ (probe temperature) as a function of the input parameter $n$, the D2 brane flux . For all simulations, $N=15$, and there are no background fluctuations.}\label{fig:fluxscan}
\end{figure}

We track the dimensionless D2 brane flux through a parameter labeled $n$. We do find a correlation between $n$ and the thermal properties of the probe as shown in Figure~\ref{fig:fluxscan}. 
We are able to identify the following scaling laws
\begin{equation}
	R_{eq} \propto n\ \ \ ,\ \ \ \tau_{th} \propto n^{-1/2}\ ;
\end{equation}
\begin{equation}
	r_{th}^2 = t_p \propto n^2\ \ \ ,\ \ \ \tau_l \propto n^{-1}\ .
\end{equation}
These results indicate a sensitivity of the thermal properties of the probe to black hole hair, if present. Larger D2 flux results in larger thermal equilibrium size of the probe, which correlates with the proximity of a larger background black hole. Larger D2 brane flux seems to also hasten the thermalization. The probe's equilibrium temperature seems however to increase with the background flux, a rather counter-intuitive scaling relation from the perspective of the naive Matrix black hole model~\cite{Horowitz:1997fr,Banks:1997tn,Banks:1997hz}.

\subsection{Results: the in-fall problem}

\begin{figure}
	\begin{center}
		\includegraphics[width=6in]{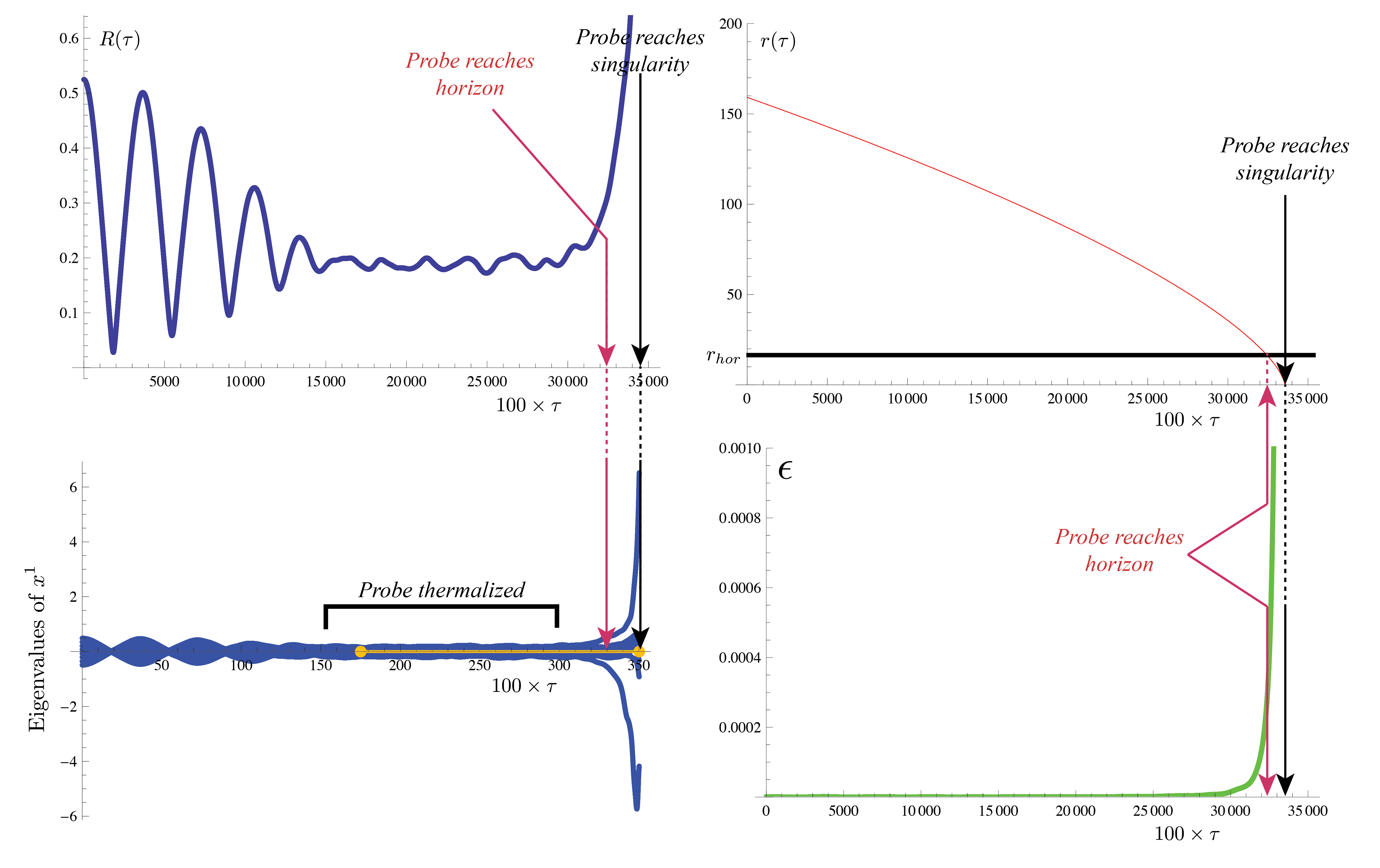} 
	\end{center}
	\caption{The probe's evolution as it falls into a Schwarzschild black hole. The black hole temperature is $t_{bh}=0.004$, corresponding to tidal force parameter $m_1=6\times10^{-7}$ as defined in the main text. The other parameters of the simulation are $N=15$ for number of D0 branes, $g_s=10^{-6}$ for the string coupling, and $r_0/r_{h}=10$. In the top two graph, we plot the probe's size and the probe's radial coordinate distance from the black hole as a function of time. In the lower left figure, we see the eigenvalue evolution and the thermalization prior to reaching the horizon. Finally, in the lower right graph, we plot the $\epsilon$ parameter, defined later on in the main text: this parameter tracks the adiabatic regime of validity of the Fermi normal coordinates. If $\epsilon \sim 1$, the simulation results cannot be trusted. As we can see, the parameter remains well within the reliable range as the probe crosses the horizon, but not much beyond.}\label{fig:bhscan}
\end{figure}

In the third and last scenario, we drop the probe D0 branes from rest near the vicinity of the Schwarzschild black hole. We start at an initial radial distance $r_0$ from the horizon $r_h$ such that $r_0/r_h=10$; and we track the evolution as a function of local Fermi normal coordinate time, as the probe crosses the horizon and is sucked into the black hole singularity. We consider a large black hole to assure that our simulation are reliable near the horizon. In units of $1/\ell$, we choose the temperature of the hole as $t=0.004$. In units of $\ell$, the flight time to the singularity is $\tau_{fl}\simeq 336$, while the time to reach the horizon is $\tau_{hor}\simeq 325$. And $\tau_{hor}$ is well within the regime of validity of our simulation. Figure~\ref{fig:bhscan} summarizes the results for a collection of $N=15$ D0 branes.
We find that the probe thermalizes due to internal dynamics as seen earlier, when it is away from the horizon. We find surprisingly that this thermalization is insensitive to the black hole temperature! The thermalization time and the equilibrium probe temperature and size do not depend on the black hole parameters. As the probe approaches the horizon, we do however see the probe explode in size. Our formalism however breaks down before the probe can reach the size of the horizon. The trend however is consistent with a probe expanding violently in size as it nears the horizon, as proposed in~\cite{Murugan:2006sn}. 

We leave the conclusions of these results to the Conclusion and Outlook Section at the end, Section~\ref{sub:conclusion}. Before that however, we present the details progressively. In Section~\ref{sub:setup}, we develop the problem and lay out the parameter space we want to explore. In Section~\ref{sub:thermalization}, the Langevin problem is formulated for our Matrix theory system, including a delicate thermal back-reaction mechanism. In Section~\ref{sub:parameters}, the regime of validity of the three cases we study is elaborated in detail. And in Section~\ref{sub:numerics}, some of the technical numerical considerations are presented.

\section{The setup}
\label{sub:setup}

Consider $N$ D0 branes in arbitrary supergravity background fields. The non-Abelian action is given  in terms of the coordinates of the D0 branes -- represented by $N\times N$ $U(N)$ matrices $\Phi^\mu$ with $\mu=1\cdots 9$
\begin{eqnarray}
	S &=& - \frac{1}{g_s l_s} \int dt\ \mbox{{\bf STr}} \left\{ e^{-\phi} \left( - \left( P \left[ E_{00} + E_{0\mu} \left( Q^{-1}-\delta \right)^{\mu\nu} E_{\nu0} \right] \right) \right)^{1/2} \left( \mbox{Det} Q \right)^{1/2}
	\right. \nonumber \\
	&+& \left.\frac{1}{g_s l_s} \int \mbox{{\bf STr}} \left\{ P \left[ e^{i\lambda i_\Phi i_\Phi} \left( \sum C^{(n)} e^B \right) \right] \right\}
	 \right\}\label{eq:myers}
\end{eqnarray}
where $\lambda = 2\pi l_s^2$, $Q^\mu_\nu=\delta^\mu_\nu+i\lambda [\Phi^\mu,\Phi^\rho] E_{\rho\nu}$, and $E_{\mu\nu} = G_{\mu\nu}+B_{\mu\nu}$ -- all given in the notation used in~\cite{Myers:1999ps}. This action is derived using T-duality symmetry. As an expansion in powers of $\lambda$, it is known to agree with direct open string computations to order $\lambda^3$~\cite{Bain:1999hu}. Our goal is to study dynamics of D0 branes in a certain restricted class of backgrounds to order $\lambda^2$. Our assumptions are as follows: 

\begin{enumerate}
	\item We focus on a non-trivial background metric $G_{\mu\nu}$ and 3-form RR potential $C^{(3)}$, a constant dilaton $e^\phi=g_s$, and set all other supergravity fields to zero. We want to analyze the evolution of a probe consisting of $N$ D0 branes falling through background fields that may represent a Schwarzschild black hole. Hence, we need to at least keep track of the background metric seen by the probe. In the Matrix black hole picture, there are hints of a D2 brane structure stretched at the would-be black hole horizon~\cite{Banks:1997tn}. Hence, we also consider a non-zero flux from $C^{(3)}$. For simplicity, we eliminate all other fields.
	\item We consider an expansion of this action to order $\lambda^2$. This results in restrictions on the size of the probe, its speed of evolution, and the background metric. We will list these restrictions later.
	\item At order $\lambda^2$, it is easy to check that the $U(1)$ sector of the coordinates $\Phi^\mu$ -- corresponding to the center of mass of the probe -- decouples from the $SU(N)$ sector. We focus on the relative dynamics of the probe D0 branes and drop the $U(1)$ center of mass dynamics. Put differently, we will track the evolution of the probe D0 branes in their center of mass frame.
	\item In the center of mass frame of the probe, we adopt Fermi normal coordinates. This means that the metric at the origin is Minkowski, its first derivatives are zero, and the second derivatives of the metric -- the tidal forces -- provide the leading gravitational effects. We also assume that the probe does not back-react onto the background, except thermally -- as we will explain in detail later.
\end{enumerate}

Under these conditions, we are then left with the Hamiltonian~\cite{Murugan:2006sn}
\begin{equation}
	H = \frac{(2\pi l_s^2)^2 }{g_s l_s}\ \mbox{{\bf Tr}}
	\left\{
	\frac{( {\dot{\Phi}}^\mu )^2}{2} + M_{\mu\nu} \Phi^\mu \Phi^\nu
	+i N_{\alpha\beta\gamma} \left[ \Phi^\alpha , \Phi^\beta \right] \Phi^\gamma
	-\frac{1}{4} \left[ \Phi^\mu , \Phi^\nu \right]^2
	 \right\}
\end{equation}
where the $\Phi^\mu$'s are $N\times N$ hermitian {\em traceless} matrices representing the non-commutative coordinates of $N$ probe D0 branes. Having chosen the static gauge for the gauge field on the world-line, we also need to supplement the equations of motion with the constraint
\begin{equation}
	\left[ \Phi^\mu, {\dot{\Phi}}_\mu \right] = 0\ .
\end{equation}
The physical position coordinates of the D0 branes in nine space directions are given by
\begin{equation}
	X^\mu = 2\pi l_s^2 \Phi^\mu\ .
\end{equation}
The background metric appears through\footnote{$M_{\mu\nu}$ also depends on D0 brane flux $F^{(2)}=dC^{(1)}$, as shown in~\cite{Sahakian:2000bg}. Hence, our analysis throughout can be thought of including the effects of D0 brane flux in the background through the consideration of the $M_{\mu\nu}$. The existence of non-zero $C^{(1)}$ however also introduces a magnetic, velocity dependent term in the Hamiltonian that we have dropped. This implies that our analysis does include the leading effect of RR $F^{(2)}$ flux for small D0 brane speeds.}
\begin{equation}\label{eq:sugra1}
	M_{\mu\nu} = -\frac{1}{4} G_{tt,\mu\nu}\ ;
\end{equation}
while the D2 brane flux appears through
\begin{equation}\label{eq:sugra2}
	N_{\alpha\beta\gamma} = \frac{1}{2} C^{(3)}_{t[\alpha\beta , \gamma]}\ .
\end{equation}
See~\cite{Sahakian:2000bg} for a more general and detailed derivation of these relations.

It is convenient to use dimensionless variables instead -- labeled as $\phi^\mu$, $m_{\mu\nu}$, $n_{\alpha\beta\gamma}$, and $\tau$ -- defined as follows
\begin{equation}
	\Phi^\mu = \frac{1}{\ell} \phi^\mu\ \ \ ,\ \ \ 
	M_{\mu\nu} = \frac{m_{\mu\nu}}{\ell^2}\ \ \ ,\ \ \ 
	N_{\alpha\beta\gamma} = \frac{n_{\alpha\beta\gamma}}{\ell}\ \ \ ,\ \ \ 
	t = \tau \ell\ ,
\end{equation}
where $\ell$ has dimension of length.
This leads to the Hamiltonian
\begin{equation}
	H = \frac{(2\pi l_s^2)^2 }{g_s l_s}\frac{1}{\ell^4}\ \mbox{{\bf Tr}}
	\left\{
	\frac{1}{2}( {\dot{\phi}}^\mu )^2 + m_{\mu\nu} \phi^\mu \phi^\nu
	+i n_{\alpha\beta\gamma} \left[ \phi^\alpha , \phi^\beta \right] \phi^\gamma
	-\frac{1}{4} \left[ \phi^\mu , \phi^\nu \right]^2
	 \right\}
\end{equation}
where derivatives are now with respect to dimensionless time $\tau$.
We will choose $\ell$ so that the scale of the energy is given by $1/\ell$
\begin{equation}
	\frac{(2\pi l_s^2)^2 }{g_s l_s}\frac{1}{\ell^4} \equiv \frac{1}{\ell}
\end{equation}
which implies
\begin{equation}
	\ell = (2\pi)^{2/3} \frac{l_s}{g_s^{1/3}}\ .
\end{equation}
Hence, all lengths/times are measured in units of $\ell$, and energies in $1/\ell$. We then can write\footnote{An alternative scaling used in the literature is given by the choice of $\ell$ such that
\begin{equation}
	\frac{(2\pi l_s^2)^2 }{g_s l_s}\frac{1}{\ell^4} \equiv \frac{1}{g_s l_s}\ ,
\end{equation}
{\em i.e} the D0 brane mass, which implies $\ell = (2\pi)^{1/2} l_s$, the string scale.
}.
\begin{eqnarray}
	H &=&  \frac{1}{\ell}\ \epsilon \nonumber \\
	&=&  \frac{1}{\ell}\ \mbox{{\bf Tr}}
	\left\{
	\frac{( {\dot{\phi}}^\mu )^2}{2} + m_{\mu\nu} \phi^\mu \phi^\nu
	+i n_{\alpha\beta\gamma} \left[ \phi^\alpha , \phi^\beta \right] \phi^\gamma
	-\frac{1}{4} \left[ \phi^\mu , \phi^\nu \right]^2
	 \right\}\label{eq:dimensionlesshamiltonian}
\end{eqnarray}	
where $\epsilon$ is dimensionless energy. 
Denoting by $[s]$ the numerical scale in the dimensionless matrices $\phi^\mu$
\begin{equation}
	\mbox{Scale of matrix entries in  } \phi^\mu \sim [s] \Rightarrow  
	\mbox{Probe size} \propto [s]\ \ell\ .
\end{equation}
Since we will need $g_s\ll 1$, this means that $\ell\gg l_s$. Hence, we are looking at large probes and small energies compared to the string scale. 

We define the effective dimensionless coupling in the theory as
\begin{equation}\label{eq:geff}
	g_{eff}^2 \equiv \frac{g_{Y}^2}{H^3} = \frac{(2\pi)^{-2} g_s l_s^{-3}}{H^3}  = \frac{1}{\epsilon^3}\ .
\end{equation}
This is the parameter that tunes the strength of the quartic and cubic terms in the Hamiltonian. For strong coupling effects, we would expect $g_{eff}^2\sim 1$. For large $N\gg 1$, we may encountered an effective coupling given by $N/\epsilon^3$. The simulations presented in this work with clear thermalization effects correspond to $g_{eff}^2>1$. As a rule of thumb, when the dimensionless scale of the size of the probe goes below $s<0.3$, we get into the strong coupling regime. Note that classically this strong coupling regime is meant as a statement about the relative importance of the non-linear (cubic and quartic) terms in the Hamiltonian.

\subsection{Regime of validity}
\label{sec:regime}

The expansion of the original Dirac-Born-Infeld (DBI) action~(\ref{eq:myers}) is valid under the conditions
\begin{equation}
	[s]\, g_s^{1/3}\ll 1\ \ \ ,\ \ \ [s]^2\ [m]\ g_s^{4/3}\ll 1\ \ \ ,\ \ \ [\dot{s}] g_s^{2/3} \ll 1\ ,
\end{equation}
where the square brackets signify `numerical scale of', $[m]$ denotes the typical scale of the background fields $m_{\mu\nu}$, and $[n]$ that of $n_{\alpha\beta\gamma}$. These three statements follow from the convergence of the expansion of the DBI. In addition, if the probe size is small relative to the length scales in the background fields
\begin{equation}
	[s]^2 [m] \ll 1\ \ \ ,\ \ \ [s]^2 [n]^2\ll 1\ ,
\end{equation}
we then expect that the background fields will evolve slowly compared to the evolution of the probe degrees of freedom. We do not impose this last restriction when studying background fluctuations, but we will need it for the remaining analysis.

As for the background fields, first the string coupling must be small so that the leading supergravity regime is valid
\begin{equation}
	g_s\ll 1\ .
\end{equation}
We also need weak curvature scales in the background to protect from excited string states
\begin{equation}
	\left[ M \right] \ll l_s^{-2} \Rightarrow \left[ m \right] g_s^{2/3} \ll 1\ ,
\end{equation}
and similarly weak fluxes for the background field $n$
\begin{equation}
	\left[ n \right] g_s^{1/3} \ll 1\ .
\end{equation}

The energy of a configuration will have the following scaling structure
\begin{equation}\label{eq:energyestimate}
	\left[\epsilon\right]\sim \left( \left[m\right]\,\left[ s \right]^2+\left[n\right]\,\left[ s \right]^3 + \left[ s \right]^4 \right)\times N^\gamma\ ,
\end{equation}
where we have written
\begin{equation}
	\left[Tr[\phi^2]\right] \sim N^\gamma
\end{equation}
with $\gamma$ being typically a number between one and three and $N$ being the number of D0 branes. For example, $\gamma=3$ for the highly ordered spherical D0 brane configuration satisfying the $SU(2)$ algebra. As the system evolves through the classical equations of motion, we expect the configuration will be attracted toward a state where all three terms in the energy expression~(\ref{eq:energyestimate}) are of the same order
\begin{equation}\label{eq:energyestimate2}
	\left[m\right]\,\left[ s \right]^2\sim\left[n\right]\,\left[ s \right]^3 \sim \left[ s \right]^4\Rightarrow
	\ \ \ \ \left[ s \right]^2\sim \left[m\right]\ \ \ ,\ \ \ \left[ s \right]\sim \left[n\right] .
\end{equation}

In summary, we have a parameter space consisting of the string coupling $g_s$, the background fields $m_{\mu\nu}$ and $n_{\alpha\beta\gamma}$, the D0 initial size scale $s$, and the number $N$ of D0 branes. When we consider background thermal fluctuations, the scales of $m_{\mu\nu}$ and $n_{\alpha\beta\gamma}$ get set through the input temperature $t$ of the heat bath as described later. The following conditions delineate this parameter space:
\begin{itemize}
	\item Small string coupling
	\begin{equation}\label{eq:smallcouplingcond}
		g_s\ll 1\ .
	\end{equation}
	
	\item Weak supergravity fields
	\begin{equation}\label{eq:weakfieldscond}
		\left[ m \right] g_s^{2/3} \ll 1\ \ \ ,\ \ \ \left[ n \right] g_s^{1/3} \ll 1\ .
	\end{equation}
	
	\item Valid DBI expansion
	\begin{equation}\label{eq:dbicond}
		[s]\, g_s^{1/3}\ll 1\ \ \ ,\ \ \ [s]^2\ [m]\ g_s^{4/3}\ll 1\ \ \ ,\ \ \ [\dot{s}] g_s^{2/3} \ll 1\ .
	\end{equation}
	
	\item Small probe (for non-fluctuating backgrounds)
	\begin{equation}\label{eq:smallprobecond}
		\left[ s \right]^2 \left[ m \right] \ll 1\ \ \ ,\ \ \ \left[ s \right]^2 \left[ n \right]^2 \ll 1 \ .
	\end{equation}
\end{itemize}

To simplify the problem further, we will look at background fields with a lot of symmetry. Even with background fluctuations, we expect that the most symmetric scenario is the closest to equilibrium. Hence, restricting from the outset to symmetric setups should not miss the final equilibrium states we are interested in. We divide the nine dimensional space in two subspaces  
\begin{equation}
	\phi^i \rightarrow x^i\ \ \ ,\ \ \ \phi^a \rightarrow y^a
\end{equation}
with $i,j,\ldots =1\cdots 3$, $a,b,\ldots =4\cdots 9$. The momentum canonical to $x^i$ is labeled $p^i$, and that to $y^a$ is labeled $q^a$. We then consider backgrounds with $SO(3)\times SO(6)$ symmetry. This means we have two background fields in $m_{\mu\nu}$
\begin{equation}
	m_{\mu\nu} = \left\{ 
		\begin{array}{ll}
			0 & \mu \neq \nu \\
			m_1 & \mu=\nu=1,2,3 \\
			m_2 & \mu=\nu= 3,\cdots, 9
		\end{array}
	\right.
\end{equation}
which we call $m_1$ and $m_2$; and one field in $n_{\alpha\beta\gamma}$ 
\begin{equation}
	n_{\alpha\beta\gamma} = \left\{
		\begin{array}{ll}
			n \epsilon_{\alpha\beta\gamma} & \alpha, \beta, \gamma = 1,2, 3 \\
			0 & \mbox{Otherwise}
		\end{array}
	\right.
\end{equation}
which we simply call $n$. In part of the discussion, we will freeze the six dimensional subspace and consider probe dynamics in $3+1$ dimensions only. In this setting, we split $m_1$ into $m_0$ and $m_1$ as follows:
\begin{equation}\label{eq:bhsubspace}
	m_{\mu\nu} = \left\{ 
		\begin{array}{ll}
			m_0 & \mu=\nu=1 \\
			m_1 & \mu=\nu=2,3 \\
			\infty & \mbox{Otherwise (dynamics frozen by hand)}
		\end{array}
	\right.
\end{equation}
for $SO(2)$ cylindrical symmetry in three dimensions. This is so as to capture the scenario of a probe falling into a black hole: the $1$ direction is the in-falling radial direction, while $2$ and $3$ are transverse to it.

For the BMN model~\cite{Berenstein:2002jq}, we note the special background field values in our notation
\begin{equation}
	m_1=m_2=\frac{1}{2}\ \ \ ,\ \ \ n = \frac{1}{2}\ .
\end{equation}
Our system however encompasses a much larger class of dynamics which we will next explore.

\section{Fluctuating backgrounds}
\label{sub:thermalization}

Part of our program is to determine the role of fluctuating background fields on the process of thermalizing probe D0 branes. These fluctuations can be quantum mechanical in nature or due to a thermal bath. Either way, the problem involves a separation of timescales: one slow time scale characterizing the probe evolution, and another fast one characterizing the background fluctuations. A key ingredient in determining the correct evolution of the probe is the effect of back-reaction: the probe changes the background fluctuations, which in turn modifies the probe's dynamics. This is a delicate mechanism that can be studied for quantum fluctuations, as well as thermal fluctuations. We focus in this work on fluctuations of thermal nature and will come to the quantum fluctuation case in a future work. To model the effect of thermal back-reaction for our system of D0 branes, we start with a simpler toy example that shares similarities to our case, and expand from there. 

\subsection{A toy example}

The process of thermalization in general relies on an interesting back-reaction mechanism. In short, a system undergoing thermalization by being in contact with a thermal reservoir is associated with two time scales, one much shorter than the other one~(see, for example~\cite{reifbook}). The shorter timescale comes from the fluctuations in the thermal bath. The longer timescale is emergent from thermalization and characterizes the evolution of the system towards thermalization. The fluctuations of the thermal bath kick the system by exchanging energy with it in such a way that the system's energy is not conserved over the longer timescales. This shifts the energy balance in the thermal bath by a relatively small amount, which however in turn can affect the bath's fluctuations significantly enough to react back on the system differently. This delicate mechanism is at the heart of thermalization in numerous physical systems and underlies for example Brownian motion dynamics. To illustrate it concretely, and in particular as it is relevant to our problem of D0 brane dynamics, consider a system consisting of a single particle of mass $m$ subject to a spring force with energy
\begin{equation}
	E = \frac{1}{2} m v^2 + \frac{1}{2} k r^2\ .
\end{equation}
In this setup however, the spring constant $k$ is not constant and is fluctuating ergodically around an average value $\overline{k}$
\begin{equation}
	 k = \langle k \rangle_{th} +\delta_0 k = \overline{k}+\delta_0 k
\end{equation}
where the $th$ subscript on the averaging indicates a thermodynamic ensemble average in the heat bath system. The thermal fluctuations $\delta_0 k$ are due to the thermal reservoir having a fixed temperature $T$ and occur on a short timescale $\delta_0 t \sim 1/T$. Note that the thermal average is then, according to the ergodic theorem of thermodynamics, the same as the time average over timescales much larger than $\delta_0 t$. These are attributes of thermodynamic equilibrium in heat baths. 

We want to understand how the system evolves under the influence of these fluctuations in $k$ and hence how the system may thermalize. Denote the longer timescale associated with the evolution of the system by $\delta t$, and $\delta t \gg \delta_0 t$ by assumption. We define a time averaging scheme for observables based on this longer timescale as
\begin{equation}
	\langle f(t) \rangle_\delta \equiv = \int_t^{t+\delta t} dt' f(t')
\end{equation}
for any function of time $f(t)$. The equations of motion will then naively become
\begin{equation}\label{eq:simpleeom}
	m\langle {\bf a}\rangle_\delta = - \langle{k\, {\bf r}}\rangle_\delta \simeq - \langle{k \rangle_{th} \, \langle {\bf r}}\rangle_\delta \simeq - \overline{k} \langle {\bf r} \rangle_\delta\ ,
\end{equation}
where the ergodic theorem and $\delta t \gg \delta_0 t$ was used.
This would be the end of the story if back-reaction is not taken into account: we say the reservoir is so big that it is unaffected by our system, and the spring constant is independently maintained at its average value $\overline k$. Our system then oscillates harmonically with average spring constant $\overline{k}$. This treatment would miss the interesting effect of thermalization that, depending on the circumstances, may affect the dynamics significantly. Instead, we need to keep track of the effect of the evolution of the system on the heat bath which then back reacts on the system. 

The fluctuations $\delta_0 k$ occur over the short timescale $\delta_0 t$ and have a prescribed distribution determined by the nature of the thermal ensemble. For example, for high enough temperatures and low enough densities, we may take the fluctuations in $k$ to obey the classical Maxwellian distribution at equilibrium
\begin{equation}\label{eq:maxwellian}
	\mbox{Prob}_{eq}(\delta_0 k) = \frac{e^{-\frac{(\delta_0 k)^2}{2 \sigma_k^2}}}{\sqrt{2\pi \sigma_k^2}}
\end{equation}
where the standard deviation $\sigma_k$ can be related to the details of the heat reservoir such as its temperature. The fluctuations in $\delta_0 k$ must {\em not} be correlated over time scales greater than $\delta_0 t$. We would expect the correlation function to decay fast as in
\begin{equation}
	\langle \delta_0 k(t)\ \delta_0 k(t+\delta t) \rangle_{th} \sim e^{-\delta t/\delta_0 t}\ .
\end{equation}
Focus on an instant in time $t'$, with $t<t'< t+\delta t$, the interval over which the position of the particle changes appreciably. To leading order, the thermal average would be
\begin{equation}
	\langle \delta_0 k(t') \rangle_{th} \simeq 0
\end{equation}
if we were to ignore back-reaction effects. Over the time interval $\delta' t = t'-t$, the system's parameters have changed by
\begin{equation}
	{\bf v}\rightarrow {\bf v}+\delta' {\bf v} \Rightarrow E \rightarrow E+\delta' E\ .
\end{equation}
In particular, $\delta' E$ is not zero since the energy exchange of our system with the heat bath would change the system's energy. In the process, the heat bath's energy changes also by an amount $-\delta' E$. Hence, we may write immediately 
\begin{equation}
	\frac{\mbox{Prob}(t',\delta_0 k)}{\mbox{Prob}(t,\delta_0 k)} = e^{-\beta \delta' E}\ ,
\end{equation}
where $\beta=1/T$ and $\mbox{Prob}(t,\delta_0 k) = \mbox{Prob}_{eq}(\delta_0 k)$ is the equilibrium probability given by~(\ref{eq:maxwellian}).
This implies that 
\begin{equation}
	\langle \delta_0 k(t') \rangle_{th} \simeq \langle \delta_0 k(t) e^{-\beta \delta' E} \rangle_{th} \simeq \langle \delta_0 k(t) (1-\beta \delta' E) \rangle_{th} = -\beta \langle \delta_0 k(t) \delta' E \rangle_{th} 
\end{equation}
for small $\beta \delta' E$.
We can easily compute $\delta' E$ as
\begin{equation}
	\delta' E = \int_{t}^{t'} dt'' \delta_0 k(t'')\ {\bf r}(t'')\cdot \frac{\delta {\bf r}(t'')}{\delta t''} = \int_{t}^{t'} dt'' \delta_0 k(t'')\ {\bf r}(t'')\cdot {\bf v}(t'') 
\end{equation}
{\em i.e.} energy is conserved if $\delta_0 k(t'')=0$. Putting things together, one then gets
\begin{eqnarray}
	\langle \delta_0 k(t') \rangle_{th} &\simeq & -\beta \int_{t}^{t'} dt'' {\bf r}(t'')\cdot {\bf v}(t'')\ \langle \delta_0 k(t)  \delta_0 k(t'') \rangle_{th} \nonumber \\
	&\simeq & -\beta\ {\bf r}(t)\cdot {\bf v}(t)\ \langle \delta_0 k(0) ^2 \rangle_{th} \delta_0 t = -\beta\ {\bf r}(t)\cdot {\bf v}(t)\ \sigma_k^2\ \delta_0 t
\end{eqnarray}
where in the last steps we used the fact that $\langle \delta_0 k(t)  \delta_0 k(t'') \rangle_{th}$ is significant only over the range $\delta_0 t$, and that ${\bf r}(t)$ and ${\bf v}(t)$ change little during this interval. Note that this expression is now non-zero and leads to a velocity dependent correction to the force law, {\em i.e.} leads to a dissipation mechanism. To see this, we look back at the equation of motion~(\ref{eq:simpleeom}), and now write
\begin{eqnarray}
	 m \langle {\bf a}\rangle_\delta &\simeq & - \langle k \rangle_{th} \, \langle {\bf r} \rangle_\delta - \int_t^{t+\delta t} dt'' \delta_0 k(t'') {\bf r}(t'') \simeq - \overline{k} \, \langle {\bf r} \rangle_\delta -\langle \delta_0 k(t) \rangle_{th} \langle {\bf r} \rangle_\delta \nonumber \\
	&=& - \overline{k} \, \langle {\bf r} \rangle_\delta + \beta\ \sigma_k^2\ \delta_0 t \langle {\bf r} \rangle_\delta \langle {\bf r} \rangle_\delta \cdot \langle {\bf v} \rangle_\delta = - \overline{k} \, \langle {\bf r} \rangle_\delta + {\bf \alpha} \cdot \langle {\bf v} \rangle_\delta \label{eq:bigeom}
\end{eqnarray}
with the tensor
\begin{equation}
	{\bf \alpha} \equiv \beta\ \sigma_k^2\ \delta_0 t \langle {\bf r} \rangle_\delta \langle {\bf r} \rangle_\delta\ .
\end{equation}
This is the Langevin equation for the system, and ${\bf \alpha}$ is the dissipation tensor. The effect of $\alpha$ is crucial in thermalizing the spring system. The timescale for thermalization is then easily identified as
\begin{equation}
	\delta t \sim \frac{|{\bf \alpha}|}{m} \gg \delta_0 t\ .
\end{equation}
This hierarchy between timescales, fast fluctuations driving a thermalization mechanism on a slower timescale, is the subtlety underlying the approach to equilibrium for most systems. 

There is a quicker way to reach at this conclusion that will be useful for us in simulating Matrix black hole dynamics later on. We may have directly jumped through the sequence of approximation by proposing that the probability distribution of the fluctuations $\delta_0 k$ is modified by the presence of the ensemble to
\begin{equation}\label{eq:modification}
	\mbox{Prob}(t,\delta_0 k) = \mbox{Prob}_{eq}(\delta_0 k)\times e^{-\beta \delta_0 E} = \mbox{Prob}_{eq}(\delta_0 k) e^{-\beta \delta_0 k\ {\bf r}\cdot {\delta {\bf r}}}
\end{equation}
where $\mbox{Prob}_{eq}(\delta_0 k)$ is the equilibrium probability distribution at temperature $1/\beta$.
For the classical limit of fluctuations which this problem is valid for, this leads to 
\begin{equation}
	\mbox{Prob}(t,\delta_0 k) \propto e^{-\frac{\delta_0 k^2}{2 \sigma_k^2}} e^{-\beta\ \delta_0 k\ {\bf r}\cdot {\delta {\bf r}}} \ .
\end{equation} 
Or equivalently the average value of the fluctuations to leading order shifts from zero to
\begin{equation}
	\langle \delta_0 k \rangle_{th} \rightarrow -\beta {\bf r}\cdot \delta {\bf r} \sigma_k^2
\end{equation}
with the effect of the thermal bath back-reacting onto the spring system included. This means that the fluctuating force on the spring is 
\begin{equation}
	- \langle \delta_0 k \rangle_{th} {\bf r} \rightarrow \beta\ {\bf r}\cdot \delta {\bf r}\ \sigma_k^2\ {\bf r} = \beta\ \sigma_k^2\ \delta_0 t\ {\bf r}\ ({\bf r}\cdot {\bf v}) \
\end{equation}
since the fluctuations occur over a timescale $\delta_0 t$. This is again what we obtained earlier in~(\ref{eq:bigeom}). The difference is that we have folded the back-reaction effect as a leading order contribution by changing the distribution of fluctuations according to the energetics of the heat bath. The advantage of using this method, as given by~(\ref{eq:modification}), will become apparent when we attempt to implement this effect in our numerical simulation. Otherwise, it would be hopelessly computationally time intensive to simulate this dynamics with large enough Matrix theory matrices.

\subsection{Thermal bath and Matrix theory}

For the case of our Matrix theory Hamiltonian given by~(\ref{eq:dimensionlesshamiltonian}), we will track fluctuations in $m_{\mu\nu}$ and $n_{\alpha\beta\gamma}$. Adopting the prescription in~(\ref{eq:modification}), we then write
\begin{equation}\label{eq:thermalshift1}
	\mbox{Prob}(t,\delta_0 m_{\mu\nu}) = \mbox{Prob}_{eq}(\delta_0 m_{\mu\nu})\times \exp({-\beta\ \delta m_{\mu\nu}\ \delta_0 \left[ {\mbox{\bf Tr}\phi^\mu \phi^\nu} \right] })
\end{equation}
for the fluctuation distribution of $m_{\mu\nu}$.  Here, $\delta_0 \left[ {\mbox{\bf Tr }\phi^\mu \phi^\nu} \right]$ measures then the change in the $D0$ brane matrix coordinates under an evolutionary time step of $\delta_0 t$. Note that $\beta$ is the inverse of the {\em dimensionless} temperature; {\em i.e}, we are writing energies and temperatures in units $1/\ell$. For $n_{\alpha\beta\gamma}$, we then have
\begin{equation}\label{eq:thermalshift2}
	\mbox{Prob}(t,\delta_0 n_{\alpha\beta\gamma}) = \mbox{Prob}_{eq}(\delta_0 n_{\alpha\beta\gamma})\times \exp ({-\beta\ \delta n_{\alpha\beta\gamma}\ \delta_0 \left[ {\mbox{STr }i \left[ \phi^\alpha, \phi^\beta \right] \phi^\gamma} \right] } )\ .
\end{equation}
The question remains as to how to determine the equilibrium distributions $\mbox{Prob}_{eq}(\delta_0 m_{\mu\nu})$ and $\mbox{Prob}_{eq}(\delta_0 n_{\alpha\beta\gamma})$ of the background fields. From~(\ref{eq:sugra1}) and~(\ref{eq:sugra2}), we know that these fluctuations correspond to massless excitations in the supergravity fields of thermal origin. We then need to look at the supergravity background to determine the energy content of a disturbance $\delta m_{\mu\nu}$ and $\delta n_{\alpha\beta\gamma}$. We look at the energy content in the supergravity part of the combined D0 branes-background system because we expect that the energy is overwhelmingly stored in the background fields. That is, we assume the probe is small enough that it can be treated as a probe of the background field configuration. 

In our conventions, the supergravity action is given by
\begin{equation}
	S = \frac{1}{2\kappa^2} \int d^{10} x\ \sqrt{-G} \left( e^{-2\phi} R - \frac{1}{4} \left( F^{(2)} \right)^2 - \frac{1}{12} e^{-2\phi} \left( H^{(3)} \right)^2 - \frac{1}{48} \left( F^{(4)} \right)^2 \right)\ ,
\end{equation}
where the RR field strengths are $F^{(2)} = dC^{(1)}$ and $F^{(4)} = dC^{(3)}$, while the NSNS field strength is $H^{(3)} = dB^{(2)}$. The dilaton background is taken constant as discussed earlier. We are interested in general scaling behavior of the fluctuations with respect to temperature and string coupling. The eventual goal is to simulate the fluctuations and study the response of the system as a function of temperature and string coupling -- as opposed to actually according physical numerical importance to, for example, the temperature value. We are hence looking for extracting the qualitative features of the distribution functions $\mbox{Prob}_{eq}(\delta_0 m_{\mu\nu})$ and $\mbox{Prob}_{eq}(\delta_0 n_{\alpha\beta\gamma})$ from supergravity. We can read off the energy content of a fluctuation by looking at the time-time component of the energy momentum tensor in the {\em Einstein frame}, which is still given by
\begin{equation}
	T_{tt} = \frac{1}{\kappa^2} \left( R_{ab} - (1/2) G_{ab} R \right)
\end{equation}
since the Einstein tensor is unchanged under frame change involving a constant dilaton, and where $2\,\kappa^2= (2\pi)^7 l_s^8$. From the equations of motion, this gives for example
\begin{equation}
	T_{tt} = \frac{1}{(2\pi)^7 l_s^8} \frac{1}{2} F_{ti}^2+\cdots
\end{equation}
for the term coming from the electric field part of the 2-form field strength. Writing the field strengths in Taylor expansion about the center of mass of the D0 branes, we would get contributions of the form
\begin{equation}
	F_{ti} \simeq \left.C_{t,i}\right|_0+\left. C_{t,ij}\right|_0 x^j \rightarrow \left. C_{t,ij}\right|_0 x^j
\end{equation}
where the linear term is irrelevant except for the center of mass motion of the configuration: this is in the $U(1)$ part of the D0 matrices, which entirely decouples from the $SU(N)$ as can be checked from the form of our action. The term $M_{\mu\nu}$ in the DBI includes contributions from $C^{(1)}_{t,\mu\nu}$ as shown in~\cite{Sahakian:2000bg}.  Hence, we have $F_{ti}\sim \delta M\times L$, where this relation is to be read as relating the scale of $F_{ti}$ to typical  scale of our background fields $\delta M_{ij}$ and the length scale $L$ over which the background fields vary.

For the energy content of graviton fluctuations, we use Wald's prescription~\cite{waldbook} for identifying energy with the graviton field
\begin{equation}
	T_{tt} = \frac{1}{\kappa^2} \left( R_{ab}^{\{2\}} - (1/2) G_{ab}^{\{2\}} R^{\{2\}} \right)
\end{equation} 
where the $\{2\}$ superscript indicate expansion of the corresponding quantity to second order in metric perturbations $h_{\mu\nu}$ about the local flat metric
\begin{equation}
	R_{ab}^{\{2\}} - (1/2) G_{ab}^{\{2\}} R^{\{2\}} \simeq h_{tt,i} h_{tt,i}+\cdots\ .
\end{equation}
Once again, Taylor expanding about the center of mass of the D0 branes, we have the identification of scales
\begin{equation}
	h_{tt,i}\simeq \left.h_{tt,c}\right|_0 + \left.h_{tt,ij}\right|_0 x^j\cdots\rightarrow \left.h_{tt,ij}\right|_0 x^j \sim \delta M\times L
\end{equation}
using~(\ref{eq:sugra1}). One then repeats this scale analysis for every supergravity field.
Putting things together, one then gets for any RR field fluctuations a relation to the associated typically energy scale $\delta E$
\begin{equation}\label{eq:thermalfluct}
	\delta E \sim \ l_s^{1-D} L^{D+2} \delta M^2 \sim \ l_s^{1-D} L^{D} \delta N^2\ ;
\end{equation}
While for NSNS fields, we get an additional factor of $g_s^2$
\begin{equation}
	\delta E \sim \frac{1}{g_s^2}\ l_s^{1-D} L^{D+2} \delta M^2 \sim \frac{1}{g_s^2}\ l_s^{1-D} L^{D} \delta N^2\ .
\end{equation}
$\delta M$ and $\delta N$ denote the scales of the background fields in~(\ref{eq:sugra1}) and~(\ref{eq:sugra2}); $L$ is the spatial size of the excitation; and $D$ is the number of non-compact dimensions, taken as nine or three. The rest of the directions of space are assumed to be compact of size set by the string scale $l_s$, most simply through toroidal compactification. 

First, note that for fixed energy scale $\delta E$, the size of fluctuations $\delta M$ and $\delta N$ is dominated by the RR fields since we also need $g_s\ll 1$. Hence, we need to consider only~(\ref{eq:thermalfluct}).
The background fields all consist of massless degrees of freedom. In a thermal state, we then expect a black body spectrum. The modes $\delta M$ and $\delta N$ are in general functions of frequency $\omega$, but they are peaked around the peak of the black body spectrum $\omega_{pk} \sim T$, with a width in frequency of the order of the temperature as well $\delta \omega\sim T$. The length scale $L$ is the characteristic scale over which our classical background fields vary. In general, an estimate of the number $n_\gamma$ of massless particles in a box of size $L$ is given by
\begin{equation}\label{eq:ngamma}
	n_\gamma \sim T^D L^D\Rightarrow L\sim n_\gamma^{1/D} \frac{1}{T}
\end{equation}
where $1/T$ is the thermal wavelength at temperature $T$. We also know that $\delta E\sim n_\gamma  \omega_{pk}\sim n_\gamma T$. One may think that perhaps we need $n_\gamma\gg 1$ to have a reliable description of the background spectrum through {\em classical} fields. This is not necessarily true. For example, one can easily describe a coherent state with less than unity occupancy using classical fields. However, in our case, we will want $n_\gamma\gg 1$: as we shall see, we will adopt a distribution of fluctuations that does not capture black body physics for low occupancy numbers; this is done for computational efficiency. In terms of dimensionless parameters, including temperature $t$ written in units of $1/\ell$, we then end up with estimates of the fluctuation sizes for the background fields as
\begin{equation}\label{eq:scaling}
	\delta m \sim g_s^{(D-1)/6}\ t^{(D+3)/2}\ n_\gamma^{-1/D}\ \ \ ,\ \ \ 
	\delta n \sim g_s^{(D-1)/6}\ t^{(D+1)/2} \ .
\end{equation}
For example, the size of the fluctuations of the electric field in a gas a photons in a black body thermal configuration in three space dimensions would scale as $T^3$. This is one power of temperature more than the standard $\delta {\bf E}^2 \sim T^4$ scaling (where ${\bf E}$ is the electric field) because our probe is sensitive to the {\em gradient} of the electric field about its center of mass. Hence we get an additional power of the thermal wavelength which goes as $1/T$. In the same way, the gravitational effects on the probe D0 branes are tidal in nature and do not involve the net gravitational force on its center of mass. The end result is summarized with the simple scaling relations shown in~(\ref{eq:scaling}). 

We take the equilibrium distribution function for the thermal background of massless fields as Gaussian
\begin{eqnarray}\label{eq:gaussian}
	\mbox{Prob}_{eq}(\delta m) &=& \frac{\exp \left[ {-\frac{\delta m^2}{2 \sigma_{m}^2}} \right]}{\sqrt{2\pi \sigma_m^2}} \nonumber \\
	\mbox{Prob}_{eq}(\delta n) &=& \frac{\exp \left[ {-\frac{\delta n^2}{2 \sigma_{n}^2}} \right]}{\sqrt{2\pi \sigma_n^2}}\ ,
\end{eqnarray}
for each mode $\delta m_{\mu\nu}$ and $\delta n_{\alpha\beta\gamma}$, with the standard deviations $\sigma_{m}^2$ and $\sigma_{n}^2$ given by~(\ref{eq:scaling})
\begin{equation}\label{eq:standarddevs}
	\sigma_{m} = g_s^{(D-1)/6}\ t^{(D+3)/2} n_\gamma^{-1/2 D}\ \ \ ,\ \ \ 
	\sigma_{n} = g_s^{(D-1)/6}\ t^{(D+1)/2}\ .
\end{equation}
We now use equality signs by absorbing any order unity numerical coefficients in our estimates into the definition of our temperature parameter $t$ and coupling $g_s$. 
There is however one additional point we need to be particularly careful about. Given that this is a fluctuation near the peak frequency $\omega_{pk} \sim T$, we need to apply the fluctuation at every time interval $1/T$ in our simulation. Hence, we can use this distribution as long as we perturb the system at a frequency near the peak of the black body spectrum.

Note that this distribution does not capture all quantum aspects of the black body distribution. To see this, the energy in, say, a background massless field $\delta m(\omega)$ at frequency $\omega$ scales as $\delta m(\omega)^2$; and this is related to the average number $n_\gamma(\omega)$ of massless particles of type $\delta m$ by
\begin{equation}
	\delta m^2(\omega) d\omega = \alpha(\omega) n_\gamma(\omega) d\omega= \alpha(\omega) \frac{1}{\exp \left[ \beta \omega \right] -1} d\omega
\end{equation} 
where $\alpha(\omega)$ is some proportionality factor scaling as $\omega^{D-1}\times \omega$. Hence, the square of the classical field measures the number of massless particles at a given frequency. The hallmark of the black body spectrum is that the fluctuations in the number of photons are not small even for $n_\gamma(\omega) \gg 1$, unlike the ideal gas distribution. This can be expressed as
\begin{equation}
	\frac{\overline{\Delta n_\gamma^2}}{\overline{n}_\gamma^2} = \frac{1}{\overline{n}_\gamma} + 1
\end{equation}
for the black body spectrum. Contrast this with classical Maxwellian distribution fluctuations which obey
\begin{equation}
	\frac{\overline{\Delta N^2}}{\overline{N}^2} = \frac{1}{\overline{N}}
\end{equation}
which goes to zero for large $N \gg 1$.
Hence, to capture the correct quantum statistics, we need to have
\begin{equation}
	\overline{\left( \delta m^2 - \overline{\delta m^2} \right)^2} = \alpha^2 \overline{\Delta n_\gamma^2} = \alpha^2 \overline{n}_\gamma \left( 1+\overline{n}_\gamma \right) = \overline{\delta m^2} \left( 1+\overline{\delta m^2} \right)
\end{equation}
for every $\omega$, and in particular for the peak $\omega_{pk} \sim T$. The gaussian distribution~(\ref{eq:gaussian}) gives instead
\begin{equation}
	\overline{\left( \delta m^2 - \overline{\delta m^2} \right)^2} = 2 \left( \overline{\delta m^2} \right)^2\ .
\end{equation}
Thus, our fluctuations do remain important for large $\overline{n}_\gamma\gg 1$ - as needed from a black body spectrum - but not for $\overline{n}_\gamma$ of order one or less. In short, treating the background fields classically, we {\em need} to assume a large condensate of massless particles within a size given classical field profiles. Our distribution however will fail to capture the statistics for low particle occupancy. This regime also corresponds to fluctuations that have a parametrically small effect on the probe D0 branes. Hence, for the purposes of tracing the thermalization evolution of our probe, the normal distribution we use is very much adequate. This trick is needed to improve the computational efficiency of our simulations. It will translate into a lower bound on temperature that we need to restrict our simulations to, as we shall see.

\section{Parameter space}
\label{sub:parameters}

In this section, we combine the general validity bounds determined in section~\ref{sec:regime} with the three particular cases we want to focus on. For each case, we want to determine the regime of the available input parameters $g_s$, $t$, $m$, $n$, $s$, and $t$ we can trust. The three cases are:

\begin{itemize}
	\item {\bf Case I:} D0 probe in background fields that average to zero, but otherwise fluctuate thermally. This scenario corresponds to the probe embedded in a dilute gas of massless supergravity fluctuations.
	\item {\bf Case II:} D0 probe in background fields that average to non-zero values, with negligible thermal fluctuations. This scenario will attempt to explore correlations of quasi-static background fields with the thermal properties of the probe.
	\item {\bf Case III:} D0 probe falling into a Schwarzschild black hole with no thermal fluctuations. This scenario aims at determining whether the approach of the probe to a black hole horizon has a special role in thermalizing the probe. In particular, we would want to see whether the probe is prevented to reach the center of the black hole through fast thermalization.
\end{itemize}

We next proceed in analyzing the `interesting' regimes in the parameter space for each of these cases: a little qualitative analytical control over the dynamics will help in zeroing onto the highlights of the numerical simulations without exploring uninteresting deserts of parameter values or physically invalid regimes.

\subsection{Case I: Zero fields with thermal fluctuations}

We take the background fields $m_1$ and $m_2$ (henceforth collectively referred to as $m$ for simplicity), and $n$ averaging to zero with Gaussian distribution given by~(\ref{eq:gaussian}). The scale of the background fields is set by the standard deviations~(\ref{eq:standarddevs}) and/or the average back-reaction shifts obtained from ~(\ref{eq:thermalshift1}) and~(\ref{eq:thermalshift2})
\begin{equation}
	\left[\delta_{back} m\right] \sim \frac{\sigma_m}{\left[ t \right]} \left[ s \right]^2 N^\gamma\ \ \ ,\ \ \ 
	\left[\delta_{back} n\right] \sim \frac{\sigma_n}{\left[ t \right]} \left[ s \right]^3 N^\gamma\ .
\end{equation}
Given that $\sigma_m$ and $\sigma_n$ are determined by the temperature of the background gas, we are thus trading the scale of $\left[ m \right]$ and $\left[ n \right]$ for the scale of the temperature $\left[ t \right]$. Hence, our parameter space in this case consists of $g_s$, $s$, $t$, and $N$. 

\begin{figure}
	\begin{center}
		\includegraphics[scale=0.4]{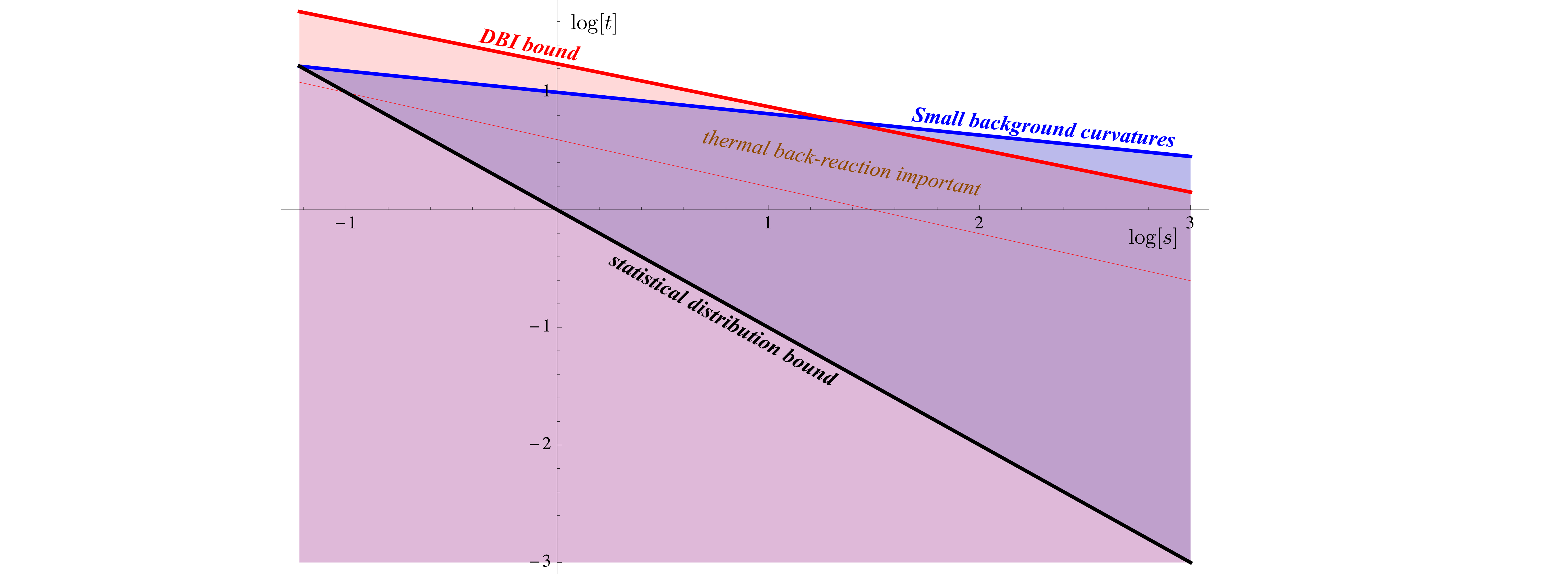} 
	\end{center}
	\caption{The parameter space delineating the regime where the simulations can be trusted for Case I. For this plot, we have taken $g_s=10^{-4}$, $\gamma=2$, $N=15$, $D=9$, and $n_{\gamma}=10^2$. There are three thick lines that bound the relevant shaded region in between: a bound arising from requiring small background curvatures (blue curve), another from convergence of the DBI action expansion (red curve), and another arising from statistical considerations (black curve, with $n_{\gamma}\sim 100$). The thin orange line bounds the region where thermal back-reaction starts becoming important. All other bounds are weaker that the ones depicted in the input parameter range we consider.}\label{fig:thermalregime}
\end{figure}

The interesting mechanism to explore in this scenario has to do with the effect of thermal back-reaction. This becomes important when
\begin{equation}\label{eq:backreactioncond}
	\sigma_m < \delta_{th} m\ \ \ ,\ \ \ \sigma_n < \delta_{th} n\ .
\end{equation}
We would like to identify the relevant values of $g_s$, $s$, $t$, and $N$ for which (1) thermal back-reaction is important {\em and} (2) the validity conditions outlined in Section~\ref{sec:regime} are satisfied. For these purposes, we can focus on fluctuations in $m$ {\em or} $n$ since they lead to the same conclusions. We know that
\begin{equation}
	\sigma_m\sim g_s^{\frac{D-1}{6}} t^{\frac{D+3}{2}}\ ,
\end{equation}
and
\begin{equation}
	\delta_{back} m \sim N^\gamma s^2 g_s^{\frac{D-1}{3}}t^{D+2}\ .
\end{equation}
Equation~(\ref{eq:backreactioncond}) then becomes
\begin{equation}\label{eq:thermalback}
	1\gg g_s> N^{-\frac{6\gamma}{D-1}} s^{-\frac{12}{D-1}} t^{-3\frac{D+1}{D-1}}
\end{equation}
where we combined the statement with the weak string coupling condition $g_s\ll 1$ from~(\ref{eq:smallcouplingcond}). The  condition of weak background fields from~(\ref{eq:weakfieldscond}) translates to 
\begin{equation}
	m\ll g_s^{-2/3}\Rightarrow 
	t\ll g_s^{-\frac{D-1}{3(D+2)}} s^{-\frac{2}{D+2}} N^{-\frac{\gamma}{D+2}}\ .
\end{equation}
While the DBI expansion gives us the conditions from~(\ref{eq:dbicond})
\begin{equation}
	[s]\, g_s^{1/3}\ll 1\ \ \ ,\ \ \ [s]^2\ [m]\ g_s^{4/3}\ll 1\ .
\end{equation}
Finally, since our statistical distribution does not capture the case of fluctuations of a small number of quanta, we get from~(\ref{eq:ngamma}) and $n_\gamma\gg 1$
\begin{equation}
	t\, s \gg 1\ .
\end{equation}
These four curves are depicted in Figure~\ref{fig:thermalregime}. We then restrict our input parameters $g_s$, $s$, $t$, and $N$ such that we are within the depicted shaded region in this Figure.

\subsection{Case II: Quasi-static backgrounds}

In the second scenario, we consider classical non-zero background field condensates, on top of which we may add {\em small} thermal fluctuations. That is, the equilibrium distributions become
\begin{eqnarray}\label{eq:gaussian2}
	\mbox{Prob}_{eq}(\delta m) &=& \frac{\exp \left[ {-\frac{(\delta m-m_c)^2}{2 \sigma_{m}^2}} \right]}{\sqrt{2\pi \sigma_m^2}} \nonumber \\
	\mbox{Prob}_{eq}(\delta n) &=& \frac{\exp \left[ {-\frac{(\delta n-n_c)^2}{2 \sigma_{n}^2}} \right]}{\sqrt{2\pi \sigma_n^2}}\ ,
\end{eqnarray}
where $m_c$ and $n_c$ are classical background field values fixed by hand. For simplicity, we also make sure that the thermal fluctuations do not overwhelm the average values of the background fields. In this case, our parameter space is given by $g_s$, $s$, $m$, $n$, and $N$. Temperature $t$ will then be inconsequential.

\begin{figure}
	\begin{center}
		\includegraphics[scale=0.4]{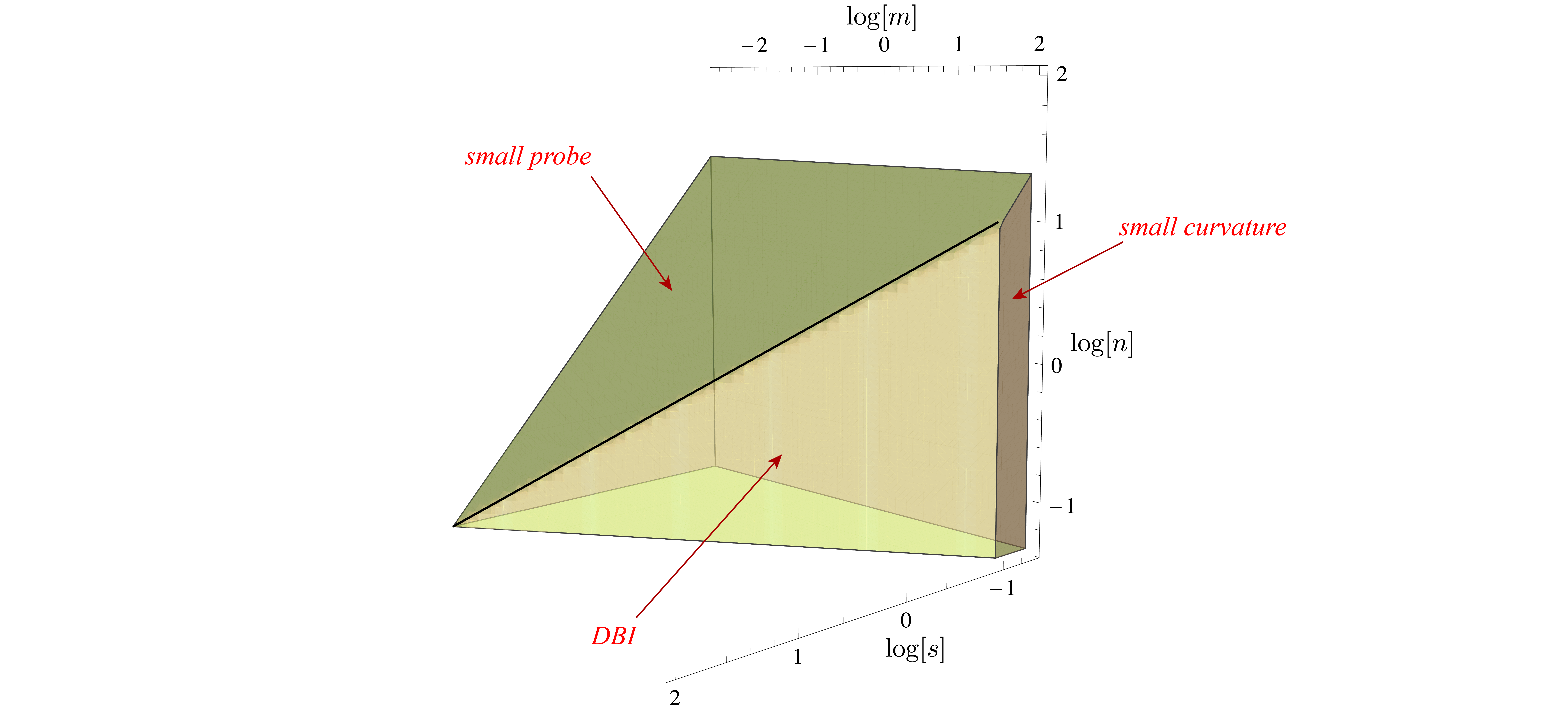} 
	\end{center}
	\caption{The region of the parameter space for which our simulations can be trusted for Case II. Three planes bound it: one arising from convergence of the DBI expansion, another from requiring small background curvatures, and another from requiring a small probe size with respect to the background length scale to assure quasi-static evolution in the center of mass frame of the probe. All other bounds are weaker that the ones depicted in the input parameter range we consider. This is a semi-infinite region, with no bounds on the smaller ends of all three axes.}\label{fig:scanregime}
\end{figure}

To assure that we are within the regime of validity of the simulations, we will need to bound ourselves to certain conditions. We need small string coupling (equation~(\ref{eq:smallcouplingcond}))
\begin{equation}
	g_s\ll 1\ ,
\end{equation}
and small background fields (equation~(\ref{eq:weakfieldscond}))
\begin{equation}
	[m] g_s^{2/3}\ll 1\ \ \ ,\ \ \ [n] g_s^{1/3} \\ 1\ .
\end{equation}
To assure that the probe motion is slow enough, we need it to be much smaller in size than the length scales set by the background fields (equation~(\ref{eq:smallprobecond}))
\begin{equation}
	[s]^2 [m] \ll 1\ \ \ ,\ \ \ [s]^2 [n]^2 \ll 1\ .
\end{equation}
Finally, to assure that the expansion of the DBI action is convergent, we need (equation~(\ref{eq:dbicond}))
\begin{equation}
	[s] g_s^{1/3} \ll 1\ \ \ ,\ \ \ [s]^2 [m] g_s^{4/3} \ll 1\ .
\end{equation}
The result of all these conditions is a region depicted in Figure~\ref{fig:scanregime}. We hence restrict again our simulations to the relevant parameter space.

\subsection{Case III: Falling into a black hole}
\label{sub:blackhole}

In this scenario, we would like to determine the time evolution of the probe D0 branes as they approach the horizon of a four dimensional Schwarzschild black hole. For simplicity, we turn off background fluctuations. In the center of mass frame of the probe, the metric looks like~\cite{fermicoordinates}
\begin{equation}
	ds^2 = - \left( 1 + \frac{G\,M}{r^3} \left( y^2+z^2-2\,x^2 \right) \right) \frac{d\tau^2}{\ell^2}+\cdots
\end{equation}
in Fermi normal coordinates. The $x$ direction is the in-fall direction, while $y$ and $z$ are transverse. The remaining six directions of space are compactified -- that is the probe's evolution is frozen in these directions.  $r$ is the location of the center of mass of the probe away from the black hole center ({\em i.e.} it is the Schwarzschild radial coordinate), and $\tau$ is the time as measured by the in-falling probe. Hence, the dynamics is $3+1$ dimensional. As the probe approaches the horizon, the metric evolves slowly compared to time scales associated with inter-probe dynamics. This is known as the adiabatic regime of the Fermi normal coordinates. The time dependence of $r$ is given by
\begin{equation}\label{eq:rt1}
	r = \frac{r_0}{2} \left( 1+\cos\omega \right)
\end{equation}
where $r_0$ is the initial radial position of the center of mass, and $\omega$ is related to the local Fermi time $\tau$ via
\begin{equation}
	d\tau^2 = \ell^2\frac{r_0}{2\,G\,M} r^2 d\omega^2\ .
\end{equation}
Or
\begin{equation}\label{eq:rt2}
	\tau = \ell\frac{r_0}{2} \sqrt{\frac{r_0}{2\, G\,M}} \left( \omega+\sin\omega \right)\ .
\end{equation}
We assume the in-fall starts with zero velocity. We then identify the background field parameters of our DBI action as
\begin{equation}
	-\frac{1}{4}G_{tt,xx} = -\frac{G\,M}{r^3}=M_0\ \ \ ,\ \ \ 
	-\frac{1}{4}G_{tt,yy} = -\frac{1}{4}G_{tt,zz} = \frac{1}{2} \frac{G\,M}{r^3} = M_1=-\frac{M_0}{2}\ ,
\end{equation}
where we have divided the three dimensional subspace into two subspaces as defined in~(\ref{eq:bhsubspace}).

The temperature of the black hole is given by
\begin{equation}
	T = \frac{1}{8\pi\,G\,M} \equiv \frac{1}{4\pi r_h}
\end{equation}
with $r_h$ being the location of the black hole horizon.
In dimensionless variables, we then have 
\begin{equation}
	m_1 = -\frac{m_0}{2} = 4\pi^2 \left( \frac{r_h}{r} \right)^3 t^2
\end{equation}
where $t$ is the black hole temperature $T$ in units of $1/\ell$.
To find $r$ as a function of time $\tau$, we start from~(\ref{eq:rt1}) and~(\ref{eq:rt2})
\begin{equation}
	8\pi t\tau \left( \frac{r_h}{r_0} \right)^{3/2} = 2 \sqrt{\frac{r}{r_0} \left( 1- \frac{r}{r_0} \right)} + \arccos \left[ \frac{2\,r}{r_0} -1 \right] 
\end{equation}
with 
\begin{equation}
	r<r_0\ .
\end{equation}
Expanding in small $r/r_0$, we get
\begin{equation}
	\frac{r}{r_0} \simeq \left( 1- 6\pi t\,\tau \left( \frac{r_h}{r_0} \right)^{3/2} \right)^{2/3}\ .
\end{equation}
The time of flight, before the probe reaches the central singularity, is given by 
\begin{equation}
	\tau_{fl} \simeq \left( \frac{r_0}{r_h} \right)^{3/2} \frac{1}{6\pi\, t}\ .
\end{equation}
We can write the background fields more conveniently as 
\begin{equation}
	m_1(\tau) = \frac{m_1(0)}{\left( 1-{3} \sqrt{m_1(0)} \tau \right)^2}
\end{equation}
with the time of flight as
\begin{equation}
	\tau_{fl} \simeq \frac{1}{3 \sqrt{m_1(0)}}
\end{equation}
where $m_1(0)$ is the initial tidal force felt by the probe at time equal to zero, away from the horizon at $r=r_0$. And a similar expression with $m_1\rightarrow -m_0/2$. The time to reach the horizon is then
\begin{equation}
	\tau_{hor} = \frac{1}{6\pi t} \left( \frac{2\pi t}{\sqrt{m_1(0)}} - 1 \right)\ .
\end{equation}

As input parameter, we get to specify the string coupling $g_s$, the initial size of the probe (through $s$), $m_1(0)$, the back hole mass or equivalently horizon $r_h$, and the initial position of the probe $r_0$. For the analysis to be reliable, we need to satisfy the following conditions.

We need small string coupling~(\ref{eq:smallcouplingcond})
\begin{equation}\label{eq:bhcond1}
	g_s\ll 1\ .
\end{equation}
And weak background fields~(\ref{eq:weakfieldscond})
\begin{equation}\label{eq:bhcond2}
	[m] g_s^{2/3}\ll 1\ .
\end{equation}
This condition needs to be satisfied at all times. This means that, given a choice of valid $m_1(0)$, the simulation is still to break down as the probe approaches the black hole singularity. Hence, we need to keep track of this condition as a function of time to know when to stop the simulation.
A proper DBI approximation requires~(\ref{eq:dbicond})
\begin{equation}\label{eq:bhcond3}
	[s] g_s^{2/3} \ll 1\ .
\end{equation}
This latter condition is now trickier to handle: in cases I and II, the probe invariably collapsed to smaller sizes; hence, assuring that $s$ is small enough at time zero was enough. In this case, we may see $s$ grow in size as a function of time, as the probe approaches the black hole. Hence, we need to check that this condition is satisfied throughout time.
Finally, these Fermi normal coordinates are reliable in the adiabatic regime. This requires the condition~\cite{fermicoordinates}
\begin{equation}\label{eq:bhcond4}
	\epsilon\equiv \frac{R^2}{r^2} \frac{r_h}{r} \ll 1\ .
\end{equation}
$R$ is the size of the probe as defined in~(\ref{eq:sizedef}). Once the probe becomes large enough to violate this condition, we need to stop the simulation.

Our strategy goes as follows. Having experimented with the previous two cases, we know that we need a time of flight of the order of $\tau_{flight}\sim 500$ with $\delta \tau = 0.01$ to maintain control over the numerics of the simulations at the level of $10\%$ error. This fixes $m_1(0) \sim 10^{-6}$. Hence, choosing any $g_s\ll 1$ is enough to handle the first two conditions~(\ref{eq:bhcond1}) and~(\ref{eq:bhcond2}). The third condition~(\ref{eq:bhcond3}) is inconsequential since we can always adjust $g_s$ to be smaller (which makes the Planck length smaller) for given $s$. The fourth condition~(\ref{eq:bhcond4}) is however one we need to keep a close eye on. It says that as the probe approaches the horizon, we need to make sure that the probe size $R$ is much less than the horizon size. This may be a problem, since one school of thought is that the probe may spread to the size of the black hole as it crosses the horizon... If this is the case, the adiabatic approximation of the Fermi coordinates breaks down and we may need to revise the setup.

\section{Numerical techniques}
\label{sub:numerics}

\begin{figure}
	\begin{center}
		\includegraphics[width=7in]{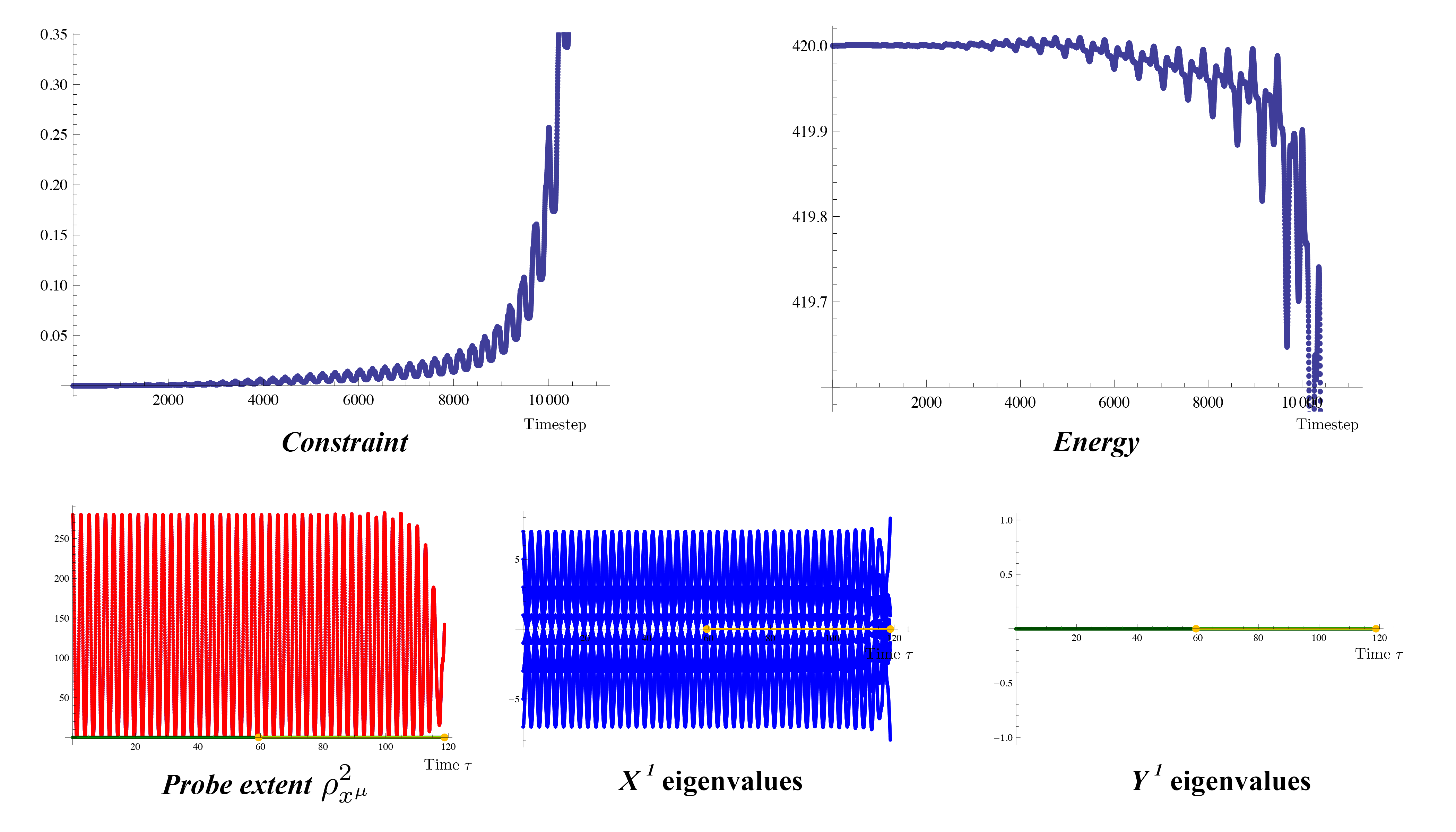} 
	\end{center}
	\caption{The time evolution of the probe when no initial fluctuations are added. We see that energy conservation and the constraint get violated by accumulated numerical errors after $40-50$ natural oscillations of the system. For the constraint, we plot the average of the norms of the matrix entries in the constraint equation.}\label{fig:stability2}
\end{figure}

The simulations were implemented as a first order coupled Hamiltonian system using a fourth-order Runge-Kutta algorithm with a typical time increment of $\delta \tau = 0.01$ -- except when the thermal fluctuation timescale is smaller $t>100$, where we reduce the size of the time-step accordingly $\delta \tau\sim 1/t$. The equations of motion are
\begin{eqnarray}
	{\dot{x}}^i &=& p^i \nonumber \\
	{\dot{y}}^a &=& q^a \nonumber \\
	{\dot{p}}^i &=& \left[ x^j, \left[ x^i, x^j \right] \right] + \left[ y^a, \left[ x^i, y^a \right] \right] - 2 m_1 x^i - 3 i n \epsilon^{ijk} \left[ x^j , x^k \right] \nonumber \\
	{\dot{q}}^a &=& \left[ y^b, \left[ y^a, y^b \right] \right] + \left[ x^i, \left[ y^a, x^i \right] \right] - 2 m_2 y^a
\end{eqnarray}
with the constraint
\begin{equation}\label{eq:simconstraint}
	[x^i,p^i]+[y^a,q^a] = 0\ .
\end{equation}
The initial conditions are set up with all the $p^i$'s, $y^a$'s, and $q^a$'s equal to zero, while the $x^i$'s satisfying the $SU(2)$ algebra
\begin{equation}
	x^i = s\,\tau^i\ \ \ ,\ \ \ \left[ \sigma^i,\sigma^j \right] = i \varepsilon^{ijk} \sigma^k
\end{equation}
in an $N\times N$ representation. In addition, small fluctuations were added to the $x^i$ matrix entries at the level of about $10\%-50\%$. Without these fluctuations, the evolution is found to be deterministic and oscillatory, as shown in Figure~\ref{fig:stability2}. 
Indeed, we use this to test the stability of our algorithm. We checked for both energy conservation and the constraint violation. This happens invariably as numerical errors accumulate over time, or when the probe starts exploding in size introducing large numbers in the simulation code that are more susceptible to computational errors. The simulation code was first implemented in Mathematica; and then, for efficiency purposes, recoded independently in objective-C. The two independent codes running on several different platforms, from Linux to Mac, were tested against each other and verified to yield the same results. Then the computation-intensive work was delegated to the objective-C version with the addition of thermal fluctuations. Single-precision floating numbers were used using the BLAS and LAPACK open source linear algebra libraries that take advantage of parallel-processing with multicore CPUs. A typical simulation for $N=15$ takes about five minutes to complete on $2-4$ cores. At larger values of $N$, we ran into memory limitations and hence an associated disk-writing time penalty. Eventually, the code will be ported to GPU systems and is expected to run up to $100$ times faster. For now, we restricted our simulations to $N<100$.

With the fluctuations added, the constraint~(\ref{eq:simconstraint}) is satisfied at the initial time. The system's evolution then assures that the constraint remains satisfied. We use the violation of the constraint to track numerical error and stop the simulation when the errors reach the $10\%$ level. 

The results shown in this work involved over $500$ simulations with varied parameters. Mathematica was used to analyze the results and generate the plots. All code is available upon request. We also have developed a visual interface front-end to the code that allows realtime exploration of D0 brane dynamics.

\section{Conclusion and outlook}
\label{sub:conclusion}

We have found that D0 brane probes have an internal mechanism, at strong coupling, for thermalizing and scrambling their initial conditions. The quartic term in the Hamiltonian is at the heart of this process. This was already determined for the BMN case in~\cite{Asplund:2011qj}. Our results extend this conclusion to a large class of backgrounds. 

This process of thermalization is a quick one: the configuration reaches an equilibrium size and temperature after a few natural oscillations. The final temperature and size, as well as the time scale of thermalization, are independent of thermal background field fluctuations -- a surprising outcome given that in most traditional statistical mechanical systems this process is intimately tied with the dynamics of the background thermal bath. Furthermore, background metric geometry on its own is not obviously correlated with this thermalization mechanism if the probe size is much smaller that the curvature length scale of the background; but some background RR fluxes do play a critical role. These observations suggest that, if a black hole geometry is to imprint its attributes onto the thermalization process of an in-falling probe, it can do so in one of two ways: (1) Have hair, perhaps along the line of the fuzzball proposal~\cite{Mathur:2005zp}. In our language, this corresponds to the effect of D2 brane flux on the thermal attributes of the probe; or (2) By expanding the in-falling probe to the order of the size of the horizon, a phenomenon we see a signature of in this work on the gauge theory side, and in previous work~\cite{Murugan:2006sn} on the gravity side. It is most likely that both mechanisms are concurrently at work in higher dimensions.

Looking more specifically at the many scaling relations we extracted from the simulations, we found that the equilibrium size of the thermalized D0 brane probe scales linearly with the number of D0 branes $N$: the more D0 branes, the bigger the equilibrium configuration, suggesting an incompressibility bound that these Planckian building blocks seem to be saturating. Furthermore, the larger this equilibrium size, the quicker the thermalization. And the equilibrium temperature of the probe increases with its equilibrium size, suggesting that perhaps we are not seeing the probe collapsing into an individual black hole -- which would have an inverse relation between its temperature and its size. Instead, the probe may be merging with a large background black hole. We see a correlation between the size of the probe and background D2 flux akin to the dielectric effect of~\cite{Myers:1999ps}, but now in this more general dynamical scenario. The Matrix black hole picture of~\cite{Horowitz:1997fr,Banks:1997tn,Banks:1997hz} has reinforced the idea that the black hole must have some order or structure, perhaps in the form of a fuzzy D-brane stretched at the horizon. While the effect of the D2 flux on the thermal properties of the probe such as its temperature is in tune with such a picture, we fail to see a signature of this order emerging within the D0 brane probe itself. Perhaps this requires a delicate phase transition within the probe~\cite{Murugan:2006sn} which can be captured when more background fluxes are turn on; in particular, D0 brane flux, which we did not consider, would give a magnetic velocity dependent force that may play a crucial role.

Intuitively, we can summarize the thermalization phenomenon we observe throughout our simulations as follows. At strong coupling, the D0 brane probe collapses in size but maintains a coherent, dense, seemingly incompressible configuration of finite smaller size. And when the system is maintained in such a confined space long enough -- a few natural oscillations of the probe -- the non-linear quartic term in the Hamiltonian scrambles the information in the probe into a thermal distribution. The process is characterized by internal attributes of the probe, or external attributes of the background that effect the equilibrium size of the probe. This latter mechanism can arise from either background fluxes, or from the probe becoming large enough to experience the IR cutoff set by the background metric curvature.

Through the works of~\cite{Sekino:2008he,HaydenPreskill:2007}, it was proposed that black holes are the fastest scramblers in the universe. The associated scrambling time $\tau_{th}$ is conjectured to scale as
\begin{equation}
	\tau_{th} t_p\sim \log N\ ,
\end{equation}
where we have written the expression in terms of our notation ($t_p$ being the probe temperature). This is to be contrasted with normal matter scramblers associated with a time scale
\begin{equation}
	\tau_{th} t_p \sim N^\alpha
\end{equation}
where $\alpha$ is a dimension dependent number. We will now argue that our results are indeed consistent with the conjectures of~\cite{Sekino:2008he,HaydenPreskill:2007}. To see this, we start from~(\ref{eq:scramblingtime}) and~(\ref{eq:scramblingtemperature}) and get
\begin{equation}\label{eq:thermal2}
	\tau_{th} t_p\sim s^3 N^2\ . 
\end{equation}
$s$ tunes the energy of the system, and hence is related to the dimensionless coupling defined in~(\ref{eq:geff}). Given we are probing a large $N$ regime, we expect that the effective coupling would instead scale as
\begin{equation}
	g_{eff}^2=\frac{N}{\epsilon^3}
\end{equation}
instead of $\epsilon^{-3}$. $\epsilon$ is the energy of the probe, and, at equilibrium, scales as $\epsilon \sim s^4 N^\gamma$ as argued in~(\ref{eq:energyestimate})-(\ref{eq:energyestimate2}). $\gamma$ is close to $2$ for a structure that is truly random. However, if there is $SU(2)$ order at equilibrium -- that is a membrane like structure to the D0 branes -- it would be close to $3$ as seen earlier for the perfectly spherical initial configuration. Looking back at~(\ref{eq:relation1}), we see that, at equilibrium, we have
\begin{equation}
	R_{eq}^2 = \frac{\mbox{\bf Tr} \left( x^\mu x^\mu\right)}{N} \sim s^2 N^2
\end{equation}
which indeed implies that the equilibrium configuration is closer to one forming D2 branes, with $\gamma = 3$. Putting things together with $\gamma=3$, we can express $s$ in terms of $g_{eff}^2=N/\epsilon^3$ in~(\ref{eq:thermal2})
\begin{equation}
	\tau_{th} t_p\sim \frac{1}{\sqrt{g_{eff}}} 
\end{equation}
with no $N$ dependence left! The stronger the effective coupling, the shorter the thermalization or scrambling time, which makes sense. And we do not have a power law dependence on the degrees of freedom $N$. The conjecture of~\cite{Sekino:2008he,HaydenPreskill:2007} suggests a $\log N$ dependence. We believe that to capture this leading behavior in $N$, we need a full quantum treatment of the problem. That all powers of $N$ cancel in this expression at the classical level is consistent with the conjecture. This also syncs well with our observation that D2 brane background flux affects the size of the configuration: if the background fields are to arise from other Matrix black hole D0 branes interacting with the probe, a D2 brane order in the background Matrix black hole would be seeding this flux that is needed to correlate thermalization properties of the probe with those of the background black hole.

For all of our simulations, a numerical instability developed after enough time had elapsed. It cannot be determined conclusively and with certainty whether this phenomenon is a real physical one or a pathology of the numerical evolution of the Hamiltonian system. For zero background fields, the system does have a flat direction that drives the system to larger entropically favored configurations -- which in turn increases numerical instability. However, we see this numerical instability even when there is no  flat direction in the system. While we determined scaling relations for the lifetime of the probe, we do not hence know whether to accord any physical significance to these.

Our technique did eventually run against computational limitations and the truly large $N$ behavior of the dynamics could not be extensively developed. Fortunately, this can easily be remedied given that our simulation code is parallelized and can be adopted relatively easily onto high performance GPU  architecture. This would take us from $4$ cores to around $400$ cores; that is a factor of $100$ decrease in computation time. We plan to implement these technical improvements in the near future.

Another direction for further exploration involves determining the effects of quantum back-reaction from the background fields onto D0 brane dynamics. It has been suggested in the fuzzball proposal context that the many fuzzball geometries are in a quantum superposition~\cite{Mathur:2008kg}; hence a probe would be sensitive to quantum fluctuations of the background geometry instead of the thermal ones investigated in this work. This topic may be accessible analytically using adiabacity techniques developed in the context of the study of Berry phases. We hope to report on this in the near future.

\section{Acknowledgments}

This work was support by NSF grant number PHY-0968726, and a grant from the Rose Hills foundation.

\bibliographystyle{utphys}

\begin{thebibliography}{10}

\bibitem{Maldacena:1997re}
J.~M. Maldacena, ``The large N limit of superconformal field theories and
  supergravity,'' {\em Adv. Theor. Math. Phys.} {\bf 2} (1998) 231--252,
  \href{http://xxx.lanl.gov/abs/hep-th/9711200}{{\tt hep-th/9711200}}.

\bibitem{Witten:1998qj}
E.~Witten, ``Anti-de Sitter space and holography,'' {\em Adv. Theor. Math.
  Phys.} {\bf 2} (1998) 253--291,
  \href{http://xxx.lanl.gov/abs/hep-th/9802150}{{\tt hep-th/9802150}}.

\bibitem{Gubser:1998bc}
S.~S. Gubser, I.~R. Klebanov, and A.~M. Polyakov, ``Gauge theory correlators
  from non-critical string theory,'' {\em Phys. Lett.} {\bf B428} (1998)
  105--114, \href{http://xxx.lanl.gov/abs/hep-th/9802109}{{\tt
  hep-th/9802109}}.

\bibitem{Banks:1996vh}
T.~Banks, W.~Fischler, S.~H. Shenker, and L.~Susskind, ``M theory as a matrix
  model: A conjecture,'' {\em Phys. Rev.} {\bf D55} (1997) 5112--5128,
  \href{http://xxx.lanl.gov/abs/hep-th/9610043}{{\tt hep-th/9610043}}.

\bibitem{Maldacena:2001kr}
J.~M. Maldacena, ``Eternal black holes in anti-de-Sitter,'' {\em JHEP} {\bf 04}
  (2003) 021, \href{http://xxx.lanl.gov/abs/hep-th/0106112}{{\tt
  hep-th/0106112}}.

\bibitem{Balasubramanian:2011ur}
V.~Balasubramanian, A.~Bernamonti, J.~de~Boer, N.~Copland, B.~Craps, {\em
  et.~al.}, ``{Holographic Thermalization},'' {\em Phys.Rev.} {\bf D84} (2011)
  026010, \href{http://xxx.lanl.gov/abs/1103.2683}{{\tt 1103.2683}}.

\bibitem{Sekino:2008he}
Y.~Sekino and L.~Susskind, ``{Fast Scramblers},'' {\em JHEP} {\bf 0810} (2008)
  065, \href{http://xxx.lanl.gov/abs/0808.2096}{{\tt 0808.2096}}.

\bibitem{HaydenPreskill:2007}
p.~Hayden and J.~Preskill, ``{Black holes as mirrors: Quantum information in random subsystems},'' {\em JHEP} {\bf 0709} (2007)
  120, \href{http://xxx.lanl.gov/abs/0708.4025}{{\tt 0708.4025}}.

\bibitem{Mathur:2010kx}
S.~D. Mathur, ``{The Information paradox and the infall problem},'' {\em
  Class.Quant.Grav.} {\bf 28} (2011) 125010,
  \href{http://xxx.lanl.gov/abs/1012.2101}{{\tt 1012.2101}}.

\bibitem{Mathur:2011uj}
S.~D. Mathur, ``{What the information paradox is {\it not}},''
  \href{http://xxx.lanl.gov/abs/1108.0302}{{\tt 1108.0302}}.

\bibitem{Mathur:2012ux}
S.~D. Mathur, ``{The information paradox: conflicts and resolutions},''
  \href{http://xxx.lanl.gov/abs/1201.2079}{{\tt 1201.2079}}.

\bibitem{Banks:1996nn}
T.~Banks, N.~Seiberg, and S.~H. Shenker, ``Branes from matrices,'' {\em Nucl.
  Phys.} {\bf B490} (1997) 91--106,
  \href{http://xxx.lanl.gov/abs/hep-th/9612157}{{\tt hep-th/9612157}}.

\bibitem{Douglas:1996yp}
M.~R. Douglas, D.~Kabat, P.~Pouliot, and S.~H. Shenker, ``D-branes and short
  distances in string theory,'' {\em Nucl. Phys.} {\bf B485} (1997) 85--127,
  \href{http://xxx.lanl.gov/abs/hep-th/9608024}{{\tt hep-th/9608024}}.

\bibitem{Mathur:2005zp}
S.~D. Mathur, ``{The Fuzzball proposal for black holes: An Elementary
  review},'' {\em Fortsch.Phys.} {\bf 53} (2005) 793--827,
  \href{http://xxx.lanl.gov/abs/hep-th/0502050}{{\tt hep-th/0502050}}.

\bibitem{Mathur:2008nj}
S.~D. Mathur, ``{Fuzzballs and the information paradox: A Summary and
  conjectures},'' \href{http://xxx.lanl.gov/abs/0810.4525}{{\tt 0810.4525}}.

\bibitem{Kaplan:2002wv}
D.~B. Kaplan, E.~Katz, and M.~Unsal, ``{Supersymmetry on a spatial lattice},''
  {\em JHEP} {\bf 0305} (2003) 037,
  \href{http://xxx.lanl.gov/abs/hep-lat/0206019}{{\tt hep-lat/0206019}}.

\bibitem{Catterall:2007fp}
S.~Catterall and T.~Wiseman, ``{Towards lattice simulation of the gauge theory
  duals to black holes and hot strings},'' {\em JHEP} {\bf 0712} (2007) 104,
  \href{http://xxx.lanl.gov/abs/0706.3518}{{\tt 0706.3518}}.

\bibitem{Anagnostopoulos:2007fw}
K.~N. Anagnostopoulos, M.~Hanada, J.~Nishimura, and S.~Takeuchi, ``{Monte Carlo
  studies of supersymmetric matrix quantum mechanics with sixteen supercharges
  at finite temperature},'' {\em Phys.Rev.Lett.} {\bf 100} (2008) 021601,
  \href{http://xxx.lanl.gov/abs/0707.4454}{{\tt 0707.4454}}.

\bibitem{Hanada:2010rg}
M.~Hanada, ``{Numerical approach to SUSY quantum mechanics and the
  gauge/gravity duality},'' \href{http://xxx.lanl.gov/abs/1011.1284}{{\tt
  1011.1284}}.

\bibitem{Myers:1999ps}
R.~C. Myers, ``Dielectric-branes,'' {\em JHEP} {\bf 12} (1999) 022,
  \href{http://xxx.lanl.gov/abs/hep-th/9910053}{{\tt hep-th/9910053}}.

\bibitem{Asplund:2011qj}
C.~Asplund, D.~Berenstein, and D.~Trancanelli, ``{Evidence for fast
  thermalization in the plane-wave matrix model},'' {\em Phys.Rev.Lett.} {\bf
  107} (2011) 171602, \href{http://xxx.lanl.gov/abs/1104.5469}{{\tt
  1104.5469}}. 5 pages, 5 figures, revtex4 format/ v2: minor typos fixed/ v3: 8
  pages, 9 figures, minor changes, includes a supplement as appeared on PRL.

\bibitem{Horowitz:1997fr}
G.~T. Horowitz and E.~J. Martinec, ``Comments on black holes in matrix
  theory,'' {\em Phys. Rev.} {\bf D57} (1998) 4935--4941,
  \href{http://xxx.lanl.gov/abs/hep-th/9710217}{{\tt hep-th/9710217}}.

\bibitem{Banks:1997tn}
T.~Banks, W.~Fischler, I.~R. Klebanov, and L.~Susskind, ``Schwarzschild black
  holes in matrix theory. ii,'' {\em JHEP} {\bf 01} (1998) 008,
  \href{http://xxx.lanl.gov/abs/hep-th/9711005}{{\tt hep-th/9711005}}.

\bibitem{Banks:1997hz}
T.~Banks, W.~Fischler, I.~R. Klebanov, and L.~Susskind, ``Schwarzschild black
  holes from matrix theory,'' {\em Phys. Rev. Lett.} {\bf 80} (1998) 226--229,
  \href{http://xxx.lanl.gov/abs/hep-th/9709091}{{\tt hep-th/9709091}}.

\bibitem{Murugan:2006sn}
A.~Murugan and V.~Sahakian, ``{Emergence of the fuzzy horizon through
  gravitational collapse},'' {\em Phys.Rev.} {\bf D74} (2006) 106010,
  \href{http://xxx.lanl.gov/abs/hep-th/0608103}{{\tt hep-th/0608103}}. 24
  pages, 4 figures/ v2: minor clarifications, citations added.

\bibitem{Bain:1999hu}
P.~Bain, ``{On the non-Abelian Born-Infeld action},''
  \href{http://xxx.lanl.gov/abs/hep-th/9909154}{{\tt hep-th/9909154}}.

\bibitem{Sahakian:2000bg}
V.~Sahakian, ``Transcribing spacetime data into matrices,'' {\em JHEP} {\bf 06}
  (2001) 037, \href{http://xxx.lanl.gov/abs/hep-th/0010237}{{\tt
  hep-th/0010237}}.

\bibitem{Berenstein:2002jq}
D.~E. Berenstein, J.~M. Maldacena, and H.~S. Nastase, ``{Strings in flat space
  and PP waves from N=4 Super Yang-Mills},'' {\em JHEP} {\bf 0204} (2002) 013,
  \href{http://xxx.lanl.gov/abs/hep-th/0202021}{{\tt hep-th/0202021}}.

\bibitem{reifbook}
F.~Reif, {\em Fundamentals of Statistical and Thermal Physics}.
\newblock Waveland Pr Inc, 2008.

\bibitem{waldbook}
R.~Wald, {\em General Relativity}.
\newblock University Of Chicago Press, 1984.

\bibitem{fermicoordinates}
M.~C.~W. Manasse, F.~K., ``Fermi normal coordinates and some basic concepts in
  differential geometry,'' {\em J. Math. Phys.} {\bf 4} ((1963);) 735.

\bibitem{Mathur:2008kg}
S.~D. Mathur, ``{Tunneling into fuzzball states},'' {\em Gen.Rel.Grav.} {\bf
  42} (2010) 113--118, \href{http://xxx.lanl.gov/abs/0805.3716}{{\tt
  0805.3716}}.

\end{thebibliography}
\providecommand{\href}[2]{#2}\begingroup\raggedright\endgroup

\end{document}